\documentclass[12pt]{article}
\usepackage{epsf,amsfonts,amssymb,epsfig,amsmath,mathtools,graphics,slashed}
\usepackage{hyperref,hep,graphicx,subfig}
\usepackage{rotating}
\usepackage{xcolor}
\usepackage{verbatim}
\usepackage[final]{showlabels}

\definecolor{greenish}{RGB}{0,190,0}

\definecolor{yellowish}{RGB}{190,190,0}

\definecolor{bluish}{RGB}{0,0,190}

\newcommand{\nn}{\notag \\}

\newcommand{\ves}{\varepsilon}

\makeatletter

\@addtoreset{equation}{section}
\makeatother

\begin{document}

\begin{titlepage}

\vfill


\vfill

\begin{center}
   \baselineskip=16pt
   {\Large\bf Nearly Critical Holographic Superfluids}
  \vskip 1.5cm
  \vskip 1.5cm
      Aristomenis Donos and Polydoros Kailidis\\
   \vskip .6cm
   \begin{small}
      \textit{Centre for Particle Theory and Department of Mathematical Sciences,\\ Durham University, Durham, DH1 3LE, U.K.}
   \end{small}\\            
\end{center}

\vfill

\begin{center}
\textbf{Abstract}
\end{center}
\begin{quote}
We study the nearly critical behaviour of holographic superfluids at finite temperature and chemical potential in their probe limit. This allows us to examine the coupled dynamics of the full complex order parameter with the charge density of the system. We derive an effective theory for the long wavelength limit of the gapless and pseudo-gapped modes by using analytic techniques in the bulk. We match our construction with Model F in the classification of Hohenberg and Halperin and compute the complex dissipative kinetic transport coefficient in terms of thermodynamics and black hole horizon data. We carry out an  analysis of the corresponding modes and argue that at finite density the dispersion relations are discontinuous between the normal and the broken phase. We compare and contrast our results with earlier numerical work.
\end{quote}

\vfill

\end{titlepage}

\setcounter{equation}{0}

\section{Introduction}
The holographic duality provides a laboratory to analyse the behaviour of large classes of strongly coupled systems \cite{Aharony:1999ti,Witten:1998qj}. In a certain large $N$ limit, large classes of field theories become dual to classical theories of gravity. Using holography as a tool kit is particularly helpful in dealing with real time physics when finite temperature and chemical potential are involved \cite{Hartnoll:2016apf}. More generally, holography is a powerful tool to study field theories deformed by relevant deformations. 

The geometries dual to the field theory thermal states are black holes of Hawking temperature equal to the field theory temperature.  In the most well understood case of conformal field theories, the bulk geometries asymptote to Anti de-Sitter space (AdS). The chemical potentials for the charges of global symmetries are fixed by the asymptotic behaviour of the gauge fields dual to the corresponding field theory Noether current operators. Likewise, the deformation parameters of other irrelevant operators are fixed by the boundary conditions of their bulk duals at the time-like conformal boundary of AdS.

In this paper we will be particularly interested in the intersection two areas that holography has already seen many applications. The first one is the study of thermal phase transitions and symmetry breaking. One of the most prominent examples is the superfluid phase transition which was pioneered in \cite{Hartnoll:2008vx,Gubser:2008px}. With applications in condensed matter physics in mind, examples where spacetime symmetries are broken were also realised holographically \cite{Nakamura:2009tf,Donos:2011bh,Donos:2011ff,Donos:2011qt}.

From the point of view of the bulk theory, continuous phase transitions are driven by perturbative instabilities which can lead to spontaneous symmetry breaking in the stable phase. In such a case, a new gapless mode appears in the theory, the dual of the Goldstone mode. At the same time, the mode which drives the transition acquires a small gap which closes to zero when approaching the critical point. This gapless collective degree of freedom is precisely the Higgs/amplitude mode. 

The second area of applied holography that we will be interested in this paper is the effective theory governing the dynamics of low energy modes close to the critical point and incorporate the almost gapless Higgs mode. In the language of superfluids, the usual description of hydrodynamics away from the critical point captures the long wavelength behaviour of the phase of the order parameter \cite{PhysRev.60.356,PhysRev.72.838,KHALATNIKOV198270,ISRAEL198179}. Our aim is to enlarge the effective theory to include fluctuations of its modulus.

Papers with similar questions have appeared in the past. However, they  either involved models which can be solved exactly close to the critical point \cite{Herzog:2010vz,Herzog:2011ec}, or numerical techniques \cite{Amado:2009ts,Amado:2013xya,Bhaseen:2012gg,Arean:2021tks}. We chose to employ analytic techniques as we want to understand the universality of the underlying physics from a boundary theory point of view. The main tool in our construction will be the techniques we have recently developed in \cite{Donos:2022www,Donos:2022uea,Donos:2022xfd,Donos:2021pkk}  to analyse dissipative effects in holographic theories. These will let us identify an equation of motion for the amplitude of the order parameter, a constitutive relation for the conserved electric current of the theory and a Josephson relation for a local chemical potential we will identify. In combination with the Ward identity for the global $U(1)$ of the theory, these will constitute a closed system of equations.

Later, we compare the resulting equations with those resulting from the Model F in the classification of Hohenberg and Halperin \cite{RevModPhys.49.435} and find exact agreement after a certain identification of the parameters in their model with our holographic results. As part of the matching procedure, we produce a holographic formula for the complex kinetic coefficient $\Gamma_0$ in terms of black hole horizon data and thermodynamic quantities of the state. It would be interesting to compare model F to holography beyond linear response.\footnote{See e.g.  \cite{Flory:2022uzp} for some recent numerical work in a direction along those lines.}.

Using our effective theory equations, we analyse the behaviour of quasinormal modes in the broken phase. By incorporating the dynamics of the almost gapless mode, we are able to commute the limits of zero gap and infinite wavelength for the fluctuations. Naively, one might expect that when holding the wavelength fixed, in the limit of small gap the modes should match with the ones of the normal phase. However, we show that this is true only at zero background chemical potential.  Moreover, we analyse interesting pole collisions in the complex plane that happen in the crossover region.

As we will see, the discontinuity mentioned above is related to the fact that the modes of oscillation of the order parameter are different between the normal and the broken phase. In the normal phase we have two copies of the same mode coming from its real and imaginary parts. However, even though the background of the broken phase is continuously connected to that of the normal phase, the mode for the fluctuations of the order parameter involve its phase in a singular manner close to the critical point. This was already evidenced from the analysis of \cite{Donos:2022xfd}, at infinite wavelengths. Interestingly,  we find that the mode for responsible charge diffusion is continuous.

Finally, we carry out a few numerical checks in order to verify some of our analytic results. In particular, we focus on reproducing the dispersion relations for the quasinormal modes that our theory predicts. The model that we chose to apply our analysis to has been studied before in \cite{Amado:2009ts} and we chose to use exactly the same set of parameters that was used there. Both our analytical and numerical results indicate that the original suggestion of \cite{Amado:2009ts} regarding the "diffusion" constant of the pseudo-diffusive mode is not accurate for small wavevectors. Interestingly, it only holds true for wavevectors of norm much larger than the gap and below any other UV scale of the theory.

Our paper is organised in six main sections. In section \ref{sec:setup} we present our holographic setup along with the necessary thermodynamics of the bulk geometries. In section \ref{sec:deriv} we employ our holographic techniques to extract all the necessary ingredients for our effective theory. In section \ref{sec:effective} we state our theory in two equivalent ways and we write the constitutive relations of the current in terms of our hydrodynamic variables. In a separate subsection, we carry out the comparison with Model F of \cite{RevModPhys.49.435}.  In section \ref{sec:modes} we examine the behaviour of the quasinormal modes of our system in various limits and point out at its discontinuities. Section \ref{sec:numerical_checks} is devoted to our numerical checks. We conclude with some discussion and conclusions in section \ref{sec:discussion}.

\section{Setup}\label{sec:setup}

Our bulk theory will have to contain a complex scalar $\psi$ which is dual to the operator $\mathcal{O}_\psi$ whose VEV will play the role of the order parameter in our system. The global $U(1)$ under which the boundary operator $\mathcal{O}_\psi$ transforms, corresponds to a local symmetry in the bulk gauged by the one-form $A_\mu$. Moreover, we will include a relevant operator $\mathcal{O}_\phi$ which will introduce an additional deformation parameter $\phi_{s}$. As we will see, the phase transition we wish to study will be driven by either tuning the chemical potential $\mu$ or the deformation parameter $\phi_{(s)}$. Our results will be valid both at finite and at zero charge density.

For our purposes, it is sufficient to consider the bulk action,
\begin{equation}\label{eq:bulk_actionv1}
S_{bulk}=\int  d^4 x \sqrt{-g}\Big(-\frac{\tau}{4}\,F^{\mu\nu}F_{\mu\nu}-\frac{1}{2} D_\mu\psi\,D^\mu\psi^\ast-\frac{1}{2}\nabla_\mu\phi\,\nabla^\mu\phi-V\Big)
\end{equation}
where $\tau$ and $V$ are in general functions of $\phi$, $|\psi|^2$. The covariant derivative acts on the complex scalar according to $D\psi=\nabla\psi+iq_e\,A\,\psi$ and the field strength of the gauge field is simply $F=dA$. For small values of our scalar fields, we will assume that the functions $\tau$ and $V$ behave according to,
\begin{align}
V&\approx \frac{1}{2}m_\psi^2\,|\psi|^2+\frac{1}{2}m_\phi^2\,\phi^2+\cdots\,,\nn
\tau&\approx 1+c_\psi\,|\psi^2|+c_\phi\,\phi+\cdots\,.
\end{align}

For the bulk geometry dual to the thermal state, we will consider a general metric which preserves the Euclidean subgroup and time translations. Without any loss of generality, this is captured by the general metric,
\begin{equation}\label{eq:background_metric}
ds^2=-U(r)dt^2+\frac{dr^2}{U(r)}+e^{2g(r)}(dx^2+dy^2)\,.
\end{equation}
One can arrive to this background in a variety of ways and the details will not be important to our analysis. As we will see, what matters is the general properties of the background geometry \eqref{eq:background_metric}.

The conformal boundary is at $r \to \infty $ and we can use the coordinate invariance of the background theory to fix the horizon $r=0$. In the asymptotic region, the functions that appear in our metric can be taken to approach,
\begin{align}\label{eq:metric_uv}
U(r)=\left(r+R\right)^2+\cdots\,,\qquad g(r)=\ln\left(r+R\right)+\cdots\,.
\end{align}
We will set the Hawking temperature of the horizon to be $T$, fixing the near horizon Taylor expansion,
\begin{align}\label{eq:metric_ir}
U(r)=4\pi T r+\cdots\,,\qquad g(r)=g^{(0)}+\cdots\,.
\end{align}

Since we will be primarily interested in the broken phase of our probe theory, the complex scalar $\psi$ will be non-trivial in the bulk geometry. In this case, the field redefinitions $\psi=\rho\, e^{iq_e\theta}$ and $B_\mu=\partial_\mu \theta+A_\mu$ are legitimate. This, brings our bulk action \eqref{eq:bulk_actionv1} to the form,
\begin{equation}\label{eq:bulk_actionv2}
S=\int d^4x  \sqrt{-g} \Big(-\frac{\tau}{4} F^2-\frac{1}{2}\nabla_\mu\rho\,\nabla^\mu\rho-\frac{1}{2}\nabla_\mu \phi\nabla^\mu\phi-\frac{1}{2}q_e^2 \rho^2 B^2-V\Big)\,,
\end{equation}
where the field strength now reads $F=dB$. The resulting equations of motions from a variation of the action \eqref{eq:bulk_actionv2} are given by,
\begin{align}\label{eq:eom}
\nabla_\mu \nabla^\mu \rho-\partial_\rho V-q^2\rho  B^2-\frac{1}{4}\partial_\rho \tau F^2=0\,,\nn
\nabla_\mu \nabla^\mu \phi-\partial_\phi V-\frac{1}{4}\partial_\phi \tau F^2=0\,,\nn
\nabla_\mu (\tau F^{\mu\nu})-q_e^2 \rho^2 B^\nu=0\,.
\end{align}

In our construction we will consider background solutions of these equations that correspond to deforming the theory by a chemical potential $\mu$ and scalar deformation parameter $\phi_{s}$. To achieve this, we will consider backgrounds with
\begin{align}\label{eq:background_probe}
\rho=\rho(r)\,,\qquad \phi=\phi(r)\,,\qquad B=B_t(r)\,dt\,.
\end{align}
Near the horizon at $r=0$, regularity imposes the expansion,
\begin{align}
\rho(r)=\rho^{(0)}+\cdots\,,\qquad
\phi(r)=\phi^{(0)}+\cdots\,,\qquad
B_t(r)=B_t^{(0)} r+\cdots\,,
\end{align}
where $\rho^{(0)}$, $\phi^{(0)}$ and $B_t^{(0)}$ are constants of integration which need to fixed.

Close to the conformal boundary at $r\to\infty$, our physical considerations suggest the power series expansions,
\begin{align}
\rho(r)&=\frac{\rho_s}{\left(r+R\right)^{3-\Delta_\psi}}+\cdots+\frac{\rho_v}{\left(r+R\right)^{\Delta_\psi}}+\cdots\,,\nn
\phi(r)&=\frac{\phi_s}{\left(r+R\right)^{3-\Delta_\phi}}+\cdots+\frac{\phi_v}{\left(r+R\right)^{\Delta_\phi}}+\cdots\,,\nn
B_t(r)&=\mu-\frac{\varrho}{r+R}+\cdots\,,
\end{align}
where the conformal dimensions $\Delta_\psi$ and $\Delta_\phi$ of the dual operators $\mathcal{O}_\psi$ and $\mathcal{O}_\phi$ are fixed by the bulk masses according to $\Delta_\psi\,(\Delta_\psi-3)=m_\psi^2$ and $\Delta_\phi\,(\Delta_\phi-3)=m_\phi^2$. For the purposes of this paper, we will be setting the complex scalar source $\rho_s$ equal to zero. In terms of the backgrounds, this will guarantee that we have no explicit breaking of the $U(1)$ symmetry. The constant of integration $\mu$ is the field theory chemical potential and $\varrho$ is the corresponding charge density. It is worth noting, that our background thermal states will eventually be parametrised by the temperature $T$, the chemical potential $\mu$ and the scalar deformation parameter $\phi_s$.

It is now worth describing the phase diagram of the system we are considering. Its precise details will depend on the parameters of our theory. For our purposes, we will be interested in the class of theories where a thermal phase transition does take place. In this case, for some fixed values of $T$ and scalar deformation $\phi_s$, there is a critical value for the chemical potential $ \mu_c(T,\phi_s)$ above which we can find  solutions with a source-free non-trivial $\rho$. These backgrounds correspond to the thermal states of the broken phase. The hypersurface $\left(T,\phi_s,\mu_c(T,\phi_s) \right)$ defines a critical surface on which the energy difference between the broken and the normal phase is exactly zero. In the following sections, we will consider phase transitions which are driven by holding the $T$ fixed and varying $\phi_s$ and $\mu$.

\subsection{Holographic Renormalisation and Thermodynamics}\label{sec:thermo}

In this subsection we will present some of the thermodynamic properties of our system that will be useful in our construction. The first step in extracting meaningful quantities form the boundary theory point of view is holographic renormalisation \cite{Skenderis:2002wp}. Equally important is the fact that holographic renormalisation  is crucial in order to make the variational problem well defined in the bulk \cite{Papadimitriou:2005ii}. In order to render the bulk action \eqref{eq:bulk_actionv1} finite, a suitable set of boundary counterterms is required. The precise form can depend on the details of the theory but a universal set of counterterms is given by,
\begin{align}\label{eq:bdy_action}
S_{bdr}=&-\frac{1}{2}\int_{\partial M}d^{3}x\,\sqrt{-\gamma}\,[(3-\Delta_{\phi})\phi^2-\frac{1}{2\Delta_{\phi}-5}\,\partial_{a}\phi\partial^{a}\phi]\notag\\
&\quad -\frac{1}{2}\int_{\partial M}d^{3}x\,\sqrt{-\gamma}\,[(3-\Delta_{\psi})|\psi|^2-\frac{1}{2\Delta_{\psi}-5}\,D_{a}\psi D^{a}\psi^{\ast}]+\cdots\,,
\end{align}
where $\gamma_{\alpha\beta}$ is the induced metric on the asymptotic hypersurface $\partial M$ of constant radial coordinate $r$. The higher order terms can include higher derivatives of the bulk fields which in our approximation become irrelevant.

In order to find the probe's contribution to the free energy of the system, one must Wick rotate to Euclidean time $\tau=i\,t$ and evaluate the total on-shell action $I_{tot}=I_b+I_{bdr}$. More precisely, since our system is infinite, one should evaluate the probe's contribution to the free energy density $w_{FE}=T\,I_{tot}$ in the $x-y$ plane. 

In this paper, we will be ignoring the backreaction of the matter fields of our probe theory \eqref{eq:bulk_actionv1} on the background geometry. This makes meaningful to keep the temperature $T$ fixed even during the real time evolution of the system. Given that we are only considering variations of the chemical potential $\delta\mu$ and the scalar deformation parameter $\delta\phi_s$, the first law for the grand canonical free energy $w_{FE}$ becomes\footnote{We should note that in the presence of persistent superfluid currents, the first law contains additional terms \cite{Herzog:2011ec,Bhattacharya:2011tra,Bhattacharya:2011eea} which we can ignore in our case.},
\begin{align}\label{eq:first_law}
\delta w_{FE}=-\varrho\,\delta\mu-\langle\mathcal{O}_\phi\rangle\,\delta\phi_s\,,
\end{align}
where $\varrho=\langle J^t\rangle$ is the charge density of the theory and $\langle\mathcal{O}_\phi\rangle$ is the VEV of the neutral scalar operator. In contrast to the electric charge, the expectation value $\langle\mathcal{O}_\phi\rangle$ is not a conserved quantity. However, from the statistical physics point of view we can still consider the variation of the free energy with respect to one of the couplings of the theory. Such variations would show up as extra terms in the first law like the last term in \eqref{eq:first_law}.

More generally, we can define the expectation value $\langle J^\mu\rangle$ of the conserved $U(1)$ current operator. For later reference, it will also be useful to define the thermodynamic susceptibilities through variations of $\varrho$ and $\langle\mathcal{O}_\phi\rangle$ as functions of $\mu$ and $\phi_{s}$,
\begin{align}\label{eq:suscept_defs}
\delta\varrho &=\chi_{QQ}\,\delta\mu+\nu_\mu\,\delta\phi_s\,,\nn
\delta\langle\mathcal{O}_\phi\rangle &=\nu_\mu\,\delta\mu+\nu_\phi\,\delta\phi_s\,.
\end{align}
Another quantity that will prove useful later is the horizon charge density,
\begin{align}\label{eq:rhoh}
\varrho_h=e^{2 g^{(0)}}\tau^{(0)}B^{(0)}_t\,.
\end{align}
By using the equation of motion for the vector field in \eqref{eq:eom}, one can show that in the normal phase, the horizon charge density is equal to the density $\varrho$. However, this is not true for the broken phase black holes since the bulk vector field becomes massive and Stokes' theorem doesn't apply.

We will follow very similar techniques to those of \cite{Donos:2022www,Donos:2022uea}. For this reason, we note that the VEVs around which we will construct our effect theory can be extracted from,
\begin{align}\label{eq:vevs_radial_der}
\langle J^{\mu}\rangle&=\lim_{r\to\infty}\frac{r^5}{\sqrt{-\gamma}}\left[ \frac{\partial\mathcal{L}}{\partial(\partial_r B_{\mu})}+\frac{\delta S_{bdr}^\prime}{\delta B_{\mu}}\right]\,,\nn
\langle \mathcal{O}_\phi\rangle&=\lim_{r\to\infty} \frac{r^3}{\sqrt{-\gamma}}\left[\frac{\partial\mathcal{L}}{\partial(\partial_r \phi)}+ \frac{\delta S_{bdr}^\prime}{\delta \phi}\right]\,,\nn
\langle \mathcal{O}_\psi\rangle&=\lim_{r\to\infty} \frac{r^3}{\sqrt{-\gamma}}\left[\frac{\partial\mathcal{L}}{\partial(\partial_r \psi^\ast)}+ \frac{\delta S_{bdr}^\prime}{\delta \psi^\ast}\right]\,,
\end{align}
where $\mathcal{L}$ is the Lagrangian density of the bulk action \eqref{eq:bulk_actionv2}. The above formulae are going to be directly useful to us when we consider the symplectic current of the theory. Finally, we note that the electric current satisfies the Ward identity,
\begin{align}\label{eq:Ward}
\nabla_a\langle J^a\rangle&=\frac{q_e}{2 i}\,\left(\lambda\,\langle \mathcal{O}_{\psi^\ast}\rangle-\lambda^\ast\,\langle \mathcal{O}_{\psi}\rangle \right)\,.
\end{align}
In the above, the parameter $\lambda$ is the source for the complex scalar operator $\mathcal{O}_\psi$.

Apart from the thermodynamic quantities defined above, we would like to define the stiffness parameter $w^{ij}$, in a very similar way it was done in \cite{Donos:2013cka,Donos:2019txg}. One can imagine, that instead of the homogeneous background that we consider in this paper, we could have a more general family of background which break translations with a characteristic spatial wavevector $k_i$. For example these broken phase backgrounds would be driven by a static mode of the form,
\begin{align}
\delta\rho=\delta\rho(r)\,\cos(k_1\,x+c_x)\,\cos(k_2\,y+c_y)\,,
\end{align}
where $c_x$ and $c_y$ are the zero modes of the Goldstone modes for translations. The corresponding backgrounds will then also be parametrised by the periods $2\pi/k_x$ and $2\pi/k_y$ and so will the free energy $w_{FE}$. Following very similar arguments with \cite{Donos:2019txg}, we can easily show the bulk expression,
\begin{align}\label{eq:gamma_def}
w^{ij}=\left.\partial_{k_i}\partial_{k_j}w_{FE}\right|_{k_i=0}=\delta^{ij}\,\int_{0}^\infty dr\,\rho^2(r)=\delta^{ij}\,\gamma\,.
\end{align}

A defining characteristic of the superfluid phase is the appearance of persistent supercurrents. The thermodynamic conjugate variables of these is the spatial components of the source for the electric currents $\delta\mu_i$, or more precisely the gauge invariant combination $\delta\mu_i+\partial_i\,\delta \theta_v$, with $\delta\theta_v$ the phase of the complex VEV $\langle\mathcal{O}_\psi\rangle$. This, can be read off from the asymptotic behaviour of the phase,
\begin{align}
\delta\theta\approx (r+R)^{2\Delta_\psi-3}\,\delta\theta_{(s)}+\cdots+\delta\theta_{(v)}+\cdots\,.
\end{align}
From the point of view of the variables we have chosen to work which, the asymptotics of the phase field are encoded in the asymptotic behaviour of the gauge invariant one-form field components along the field theory directions according to,
\begin{align}\label{eq:uv_bexp}
\delta B_\alpha=\frac{\partial_\alpha \delta\theta_{(s)}}{(r+R)^{3-2\,\Delta_\rho}}+\cdots+ \delta m_\alpha +\cdots+\frac{ \delta j_\alpha}{r+R}+\cdots\,,
\end{align}
where $m_{\alpha}=\partial_{\alpha}\theta_{(v)}+\delta \mu_{\alpha}$ is the gauge invariant combination of the sources.

The final thermodynamic quantity we would now like to discuss it the current susceptibility $\chi_{JJ}$. If we wanted to consider all possible thermal states of our superfluids, we would have to include backgrounds in \eqref{eq:background_probe} which contain these supercurrents. In the present work, we wish to study the effective theory around states with zero supercurrents. However, these supercurrents will be relevant to our us from a perturbative point of view, as they will be involved in the hydrodynamic modes we will consider. From the bulk point of view, our broken phase backgrounds admit a non-trivial perturbation for the bulk one-form field of the form,
\begin{equation}\label{ax}
 \delta B^i=\delta B_i^{i}(r)dx^i
\end{equation}
which behaves near the boundary behaves as,
\begin{equation}
\delta B_i^{i}=\delta m_i-\frac{\chi_{JJ}\, \delta m_i}{r+R}+\cdots\,,
\end{equation}
and $\chi_{JJ}$ is precisely the current susceptibility. Near the horizon, regularity imposes the behaviour,
\begin{equation}
\delta B_i^{i}=a_J^{(0)}\delta m_i+\cdots\,.
\end{equation}
Is is useful to note that given the perturbation $\delta B_i^{i}$, the equation of motion for the gauge field gives the relation,
\begin{align}
\chi_{JJ}\,\delta m_i=q_e^2\,\int_0^\infty dr\,\rho^2\,\delta B_i^i\,.
\end{align}
Given the fact that close to the phase transition we have approximately $\delta B_i^i\approx \delta m_i$ everywhere in the bulk, using the above equation it is easy to argue that close to the phase transition we must have,
\begin{align}\label{eq:gamma_chi}
\gamma=\frac{\chi_{JJ}}{q^2_e}+\cdots\,.
\end{align}
We may have used holography to show this relation but the deeper reason is gauge invariance with respect to the external source for the current.

\section{Extracting the effective theory}\label{sec:deriv}

In this section we will extract all the necessary ingredients to construct our effective theory. In order to achieve this, we will follow a combination of techniques and arguments developed in \cite{Donos:2022www,Donos:2022uea,Donos:2022xfd,Donos:2021pkk}.

\subsection{Expansions near the critical point}

An important ingredient of our construction, will be the expansions of the background solutions \eqref{eq:background_probe} around the critical point at $(\mu_c(\phi_s),\phi_s,T)$. At exactly the critical point, the perturbative equation of motion for the amplitude $\rho$ in \eqref{eq:eom} admits a static solution $\delta\rho_{\ast (0)}$. We will denote the background field at that point by $B_t=a_c$ and $\phi=\phi_c$.

In order to establish our notation, we are moving away from the critical point according to,
\begin{align}\label{eq:variation_broken}
\mu(\ves)=\mu_c(\phi_s,T)+\frac{\ves^2}{2}\delta\mu_{\star (2)}+\cdots\,,\nn
\phi_s(\ves)=\phi_s+\frac{\ves^2}{2}\delta\phi_{s\star (2)}+\cdots\,,
\end{align}
with $\ves$ a parametrically small number. In this notation, the parameters, $\delta\mu_{\star (2)}$ and $\delta\phi_{s\star (2)}$, define the direction that we move away from the critical point in the space of thermodynamic variables. At the same time, the background will have to change with $\ves$ accordingly. By expanding the equations of motion \eqref{eq:eom}, we can establish that the correction for the background will admit an $\ves$ expansion of the form,
\begin{align}\label{backbp}
\rho=\ves\delta\rho_{\star( 0)}+\frac{\ves^3}{3!}\delta\rho_{\star (2)}+\cdots\,,\nn
\phi=\phi_c+\frac{\ves^2}{2!}\delta \phi_{\star (2)}+\cdots\,,\nn
B_t=a_c+\frac{\ves^2}{2!}\delta a_{\star (2)}+\cdots\,,
\end{align}
along the broken phase. Following the steps of \cite{Donos:2022xfd}, it will also be useful to consider the expansion of our backgrounds along the normal phase as well. The notation we will use for this case is,
\begin{align}\label{eq:variation_normal}
\mu(\ves)&=\mu_c(\phi_s,T)+\frac{\ves^2}{2}\delta\mu_{\# (2)}+\cdots\,,\nn
\phi_s(\ves)&=\phi_s+\frac{\ves^2}{2}\delta\phi_{s\# (2)}+\cdots\,,
\end{align}
with the corresponding expansion for the normal phase backgrounds,
\begin{align}\label{eq:exp_normal}
\rho&=0\,,\nn
\phi&=\phi_c+\frac{\ves^2}{2!}\delta \phi_{\# (2)}+\cdots\,,\nn
B_t&=a_c+\frac{\ves^2}{2!}\delta a_{\# (2)}+\cdots\,.
\end{align}

From the point of view of the boundary theory, it is the asymptotic behaviour of the corrections that will be important. For the broken phase backgrounds we can write, 
\begin{align}
\delta\rho_{\star( 0)}&=\frac{\delta\rho_{\star( 0)}^v}{\left(r+R\right)^{\Delta_\psi}}+\cdots\,,\nn
\delta\phi_{\star (2)}&=\frac{\delta\phi_{s\star (2)}}{\left(r+R\right)^{3-\Delta_\phi}}+\cdots+\frac{\delta\phi_{v \star (2)}}{\left(r+R\right)^{\Delta_\phi}}+\cdots\,,\nn
\delta a_{\star (2)}&=\delta\mu_{\star (2)}-\frac{\delta\varrho_{\star(2)}}{r+R}+\cdots\,.
\end{align}
In our analysis, we will also need to behaviour of these perturbations close to the black hole horizon at $r=0$ which reads,
\begin{align}
\delta a_{\star (2)}=\delta a_{\star (2)}^{(1)}r+\cdots\,,\nn
\delta\phi_{\star (2)}=\delta\phi_{\star (2)}^{(0)}+\cdots\,,\nn
\delta\rho_{\star( 0)}=\delta\rho_{\star( 0)}^{(0)}+\cdots\,.
\end{align}
For the normal phase expansion under the variations \eqref{eq:variation_normal}, we can write very similar expressions for both the asymptotic and the near horizon expansions.

By using the definitions the thermodynamic susceptibilities of equation \eqref{eq:suscept_defs} we can write the following relations,
\begin{align}
\delta\varrho_{\star(2)}&=\chi^{\star}_{QQ}\,\delta\mu_{\star (2)}+\nu^{\star}_\mu\,\delta\phi_{s\star (2)}\nn
\delta\langle\mathcal{O}\rangle_{\star(2)}&=\nu^{\star}_\mu\,\delta\mu_{\star (2)}+\nu^{\star}_\phi\,\delta\phi_{s\star (2)}\nn
\end{align}
for the broken phase. For the normal phase expansion we can write the corresponding relations,
\begin{align}\label{eq:suscepts_norm}
\delta\varrho_{\#(2)}=\chi^{\#}_{QQ}\,\delta\mu_{\# (2)}+\nu^{\#}_\mu\,\delta\phi_{s\# (2)}\,,\nn
\delta\langle\mathcal{O}\rangle_{\#(2)}=\nu^{\#}_\mu\,\delta\mu_{\# (2)}+\nu^{\#}_\phi\,\delta\phi_{s\# (2)}\,.
\end{align}
In the hydrodynamic limit we can write the expressions,
\begin{align}
\delta\langle\mathcal{O}_\phi\rangle_{\star(2)}=\left(2\,\Delta_\phi-3\right)\,\delta\phi_{v \star (2)}\,,\qquad \delta\langle\mathcal{O}_\phi\rangle_{\#(2)}=\left(2\,\Delta_\phi-3\right)\,\delta\phi_{v \# (2)}\,,
\end{align}
but for us, it is the asymptotic form of the symplectic current that will play a direct role in our analysis. Finally, it is worth noting for the normal phase we have the additional relation $\delta\varrho_{\#(2)}=e^{2 g^{(0)}}\tau^{(0)}\delta a_{\# (2)}^{(1)}$. This is nothing but our earlier statement that in the normal phase the horizon charge density \eqref{eq:rhoh} is equal to the field theory one.

Another important part of our strategy is the set of static perturbations we will use in the Crnkovic-Witten symplectic current. Similarly to \cite{Donos:2022xfd}, the first static perturbation we will need is can be obtained from the broken phase background expansion \eqref{backbp}. Taking a derivative with respect to $\ves$ we find the static perturbation,
\begin{align}\label{btbp}
\delta\rho^{\star}&=\delta\rho_{\star (0)}+\frac{\ves^2}{2!}\delta\rho_{\star (2)}+\cdots\,,\nn
\delta\phi^\star&=\ves\,\delta\phi_{\star (2)}+\cdots\,,\nn
\delta B_t^{\star}&=\ves\,\delta a_{\star (2)}+\cdots\,.
\end{align}
This fluctuation will play a double role in our construction. The first one is as we described above, it will be used as one of the two solutions in the symplectic current. The second role is that it will be used to construct the next to leading order hydrodynamic perturbation we wish to study with our effective theory. For the same reason, we will also consider the expansion along the normal phase \eqref{eq:exp_normal} and take a derivative with respect to $\ves$ to find the perturbation,
\begin{align}\label{btnp}
\delta\rho^{\#}&=0\,,\nn
\delta\phi^\#&=\ves\,\delta\phi_{\# (2)}+\cdots\,,\nn
\delta B_t^{\#}&=\ves\,\delta a_{\# (2)}+\cdots\,.
\end{align}

The second static solution that we will use in the symplectic current is the perturbation for the supercurrent of equation \eqref{ax}. Since our effort is to extract information infinitesimally close to the critical point, we will also need to consider the $\ves$ expansion of that as well,
\begin{align}
\delta B_i^{i}=\delta B_{i(0)}^{i}+\ves^2\delta B_{i(2)}^{i}+\cdots\nn
\chi_{jj}=\ves^2 \chi_{jj (2)}+\cdots\nn
a_J^{(0)}=1+\ves^2 a_{J(2)}^{(0)}+\cdots
\end{align}
where the zeroth order solution is simply $\delta B_{i(0)}^{i}=\delta m_i$, as can be seen by solving the one-form field equation of motion \eqref{eq:eom} in the normal phase with $\rho=0$. This is exactly the argument we used to show the relation \eqref{eq:gamma_chi}.

\subsection{Hydrodynamic Perturbations}

Before specialising to the low frequency, long wavelength fluctuations we are interested in, it is worth setting up the problem for a generic perturbation that depends on the field theory coordinates. By exploiting the translations in space and time, we will assume a Fourier decomposition of the form,
\begin{equation}
\delta \mathcal{F}(r,t,x)=e^{-i\omega (t+S(r))+i \ves q x}\delta f(r)\,,
\end{equation}
for bulk fields. The function $S(r)$ a function which behaves near the horizon as $S(r)=\frac{\ln r}{4\pi T}+\cdots$ and vanishes sufficiently fast at the boundary.  We will consider the quasinormal modes in the longitudinal sector and we will ignore the y component of the gauge field as it is decoupled from the rest.
Our goal is  to describe the system near the critical point and for this reason we take the momentum to be of order $\ves q \sim \sqrt{\delta \mu},\sqrt{\delta\phi_s}$.
Given the above behaviour for the function $S$ near the horizon, imposing regular ingoing boundary conditions near the horizon leads to the expansions,
\begin{align}\label{reg}
\delta \rho(r)&=\delta \rho^{(0)}+\cdots\,,\nn
\delta \phi(r)&=\delta \phi^{(0)}+\cdots\,,\nn
\delta b_x(r)&=\delta b_x^{(0)}+\cdots\,,\nn
\delta b_t(r)&=\delta b_t^{(0)}+\cdots\,,\nn
\delta b_r(r)&=\frac{\delta b_t^{(0)}}{ 4\pi T r}+\cdots\,.
\end{align}

By following general arguments, the generic expansion of our fluctuations close to the conformal boundary is,
\begin{align}
\delta \rho(r)&=\frac{\delta\rho_s}{\left(r+R\right)^{3-\Delta_\psi}}+\cdots+\frac{\delta \rho_v}{\left(r+R\right)^{\Delta_\psi}}+\cdots\,,\nn
\delta \phi(r)&=\frac{\delta\phi_s}{\left(r+R\right)^{3-\Delta_\phi}}+\cdots+\frac{\delta \phi_v}{\left(r+R\right)^{\Delta_\phi}}\,,\nn
\delta b_x(r)&=\left(\delta\mu_x+i\, \ves\, q\, \delta \theta_v\right)+\frac{\delta j_x}{r+R}+\cdots\,,\nn
\delta b_t(r)&=\left(\delta\mu_t-i\, \omega\,  \delta \theta_v\right)+\frac{\delta j_t}{r+R}+\cdots\,.\nn
\end{align}
However, for the purposes of this paper we will be interested in the source free dynamics of the low energy modes. For this reason, we will aim to set the scalar and current sources $\delta\phi_s$, $\delta\rho_s$, $\delta\mu_a$ equal to zero.

By following arguments very similar to \cite{Donos:2022xfd}, we can show that the degrees of freedom we wish to describe are captured by the expansion,
\begin{align}
\omega&=\ves^2\,\omega_{[2]}+\cdots\,,\nn
\delta \rho&=\delta \tilde{\rho}_{(0)}+\ves\, \delta  \tilde{\rho}_{(1)}+\frac{\ves^2} {2}\,\delta  \tilde{\rho}_{(2)}\cdots\,,\nn
\delta \phi&=\ves\,\delta\tilde{\phi}_{(2)}+\cdots\,,\nn
\delta b_t&=\ves\, \delta\tilde{ a}_{(2)}+\cdots\,,\nn
\delta b_x&= \delta\tilde{ b}_{x(0)}+\ves\, \delta\tilde{ b}_{x(1)}+\ves^2\, \delta\tilde{ b}_{x(2)}+\cdots\,,\nn
\delta \theta_v&=\frac{1}{\ves}\,\delta\tilde{\theta}_{v}+\delta \tilde{\theta}_{v(0)}+\cdots\,.
\end{align}

By expanding the equations of motion in $\ves$, we can see that the only solution regular at the horizon for  $ \delta\tilde{ b}_{x(0)}, \delta\tilde{ b}_{x(1)}$ is just a constant. Moreover, the equations of motion for the fields $\{\delta \tilde{\rho}_{(0)},\delta \tilde{a}_{(2)},\delta\tilde{\phi}_{(2)}\}$ are solved by a linear combination of the following solutions $\{\delta \tilde{\rho}_{(0)}=0,\delta \tilde{a}_{(2)} = \delta a_{\#(2)},\delta\tilde{\phi}_{(2)}=\delta \phi_{\#(2)}\}$ and $\{\delta \tilde{\rho}_{(0)}=\delta \rho_{\star(0)},\delta \tilde{a}_{(2)} = \delta a_{\star(2)},\delta\tilde{\phi}_{(2)}=\delta \phi_{\star(2)}\}$. Finally, note that we can add a constant (everywhere in the bulk) to $\delta \tilde{a}_{(2)}$, which we will call for convenience $-i \omega_{[2]}\delta \theta_0$, and get another solution. As a result we can write,
\begin{align}
\delta \tilde{\rho}_{(0)}&=\delta a\,\delta\rho_{\star (0)}\,,\nn
 \delta\tilde{ a}_{(2)}&=\delta a\,\delta a_{\star (2)}-\delta a_{\# (2)}-i\omega_{[2]}\delta \theta_0\,,\nn
 \delta\tilde{ \phi}_{(2)}&=\delta a\,\delta \phi_{\star (2)}-\delta \phi_{\# (2)}\,,\nn
\delta\tilde{b}_{x(0)}&=iq\,\delta\tilde{\theta}_{v}\,.
\end{align}
An important point is that the parameters $\delta\mu_{\ast(2)}$ and $\delta\phi_{\ast(2)}$ defining the broken phase variation in the above equations are identical to the variations of the background. This can be easily seen by inspection of the equations of motion. However, the variations $\delta\phi_{s\#(2)}$ and $\delta\mu_{\#(2)}$ are left undetermined by simply looking at the equations of motion.

Regarding the neutral scalar, the net deformation for the hydrodynamic perturbation is,
\begin{align}
\delta\tilde{\phi}_{s(2)}=\delta a\,\delta\phi_{s\ast{(2)}}-\delta\phi_{s\#{(2)}}\,.
\end{align}
The requirement for zero total deformations suggests that $\delta\phi_{s\#{(2)}}$ should be such that $\delta\tilde{\phi}_{s(2)}=0$. Finally, the variation parameter $\delta\mu_{\#(2)}$ will be determined by imposing the electric current conservation Ward identity \eqref{eq:Ward}. Note that the Ward identity does not provide any non-trivial information about the static perturbations of the backgrounds. However, it is going to be essential in fixing $\delta\mu_{\#(2)}$.

For the electric current of the theory, the electric current follows the $\ves$ expansion,
\begin{align}
\delta j_t&=\ves\,\delta j_{t[1]}+\cdots\,,\nn
\delta j_x&=\ves^2\,\delta j_{x[2]}+\cdots\,,
\end{align}
which is compatible with the expansion of the vector field in equation \eqref{ids}. The above to the identifications of the asymptotic data,
\begin{align}\label{ids}
-i\omega_{[2]}\delta\tilde{\theta}_{v}&=\delta a\,\delta\mu_{\star (2)}-\delta\mu_{\# (2)}-i\omega_{[2]}\delta\theta_0\nn
0&=\delta a\,\delta \phi_{s \star (2)}-\delta\phi_{s\# (2)}\nn
\delta j_{t[1]}&=-\delta a\,\delta \varrho_{\star(2)}+\delta\varrho_{\#(2)}\nn
\delta \tilde{\phi}_{v (2)}&=\delta a\, \delta \phi_{v\star (2) }-\delta \phi_{v\# (2) }\,.
\end{align}

At the same time, the above lead to the near horizon behaviour for the time component of the on-form field,
\begin{equation}\label{bt1nh}
 \delta\tilde{ a}_{(2)}=-i\omega_{[2]}\delta \theta_0+(\delta a\,\delta a_{\star (2)}^{(1)}-\delta a_{\# (2)}^{(1)}) \,r+\cdots\,.
\end{equation}
The radial component of the one-form field admits the $\ves$ expansion,
\begin{align}
\delta b_r=\ves \delta\tilde{ b}_{r(1)}+\frac{\ves^2}{2}\delta \tilde{b}_{r(2)}+\cdots\,.
\end{align}
Near the horizon, the equation of motion \eqref{eq:eom} and regularity \eqref{reg} yields the constraint,
\begin{gather}\label{eqth}
i \omega_{[2]}\left(q^2 \tau^{(0)}_c e^{-2g^{(0)}}+q_e^2\left(\delta \rho^{(0)}_{\star (0)}\right)^2\right)\delta\theta_0=i\omega_{[2]}\,\delta \tilde{\varrho}_{h(2)}e^{-2g^{(0)}}+i\tau^{(0)}_c q^2\omega_{[2]}e^{-2g^{(0)}}\delta\tilde{ \theta}_v\,.
\end{gather}
As we will later see, this is the first equation that will be part of our effective theory and it will play the role of a Josephson relation.

\subsection{Symplectic current}
In this section we will combine the analysis we have discussed so far with the techniques developed in \cite{Donos:2022www,Donos:2022uea,Donos:2022xfd,Donos:2021pkk}. For the theory described by the bulk action \eqref{eq:bulk_actionv2}, the symplectic current density is given by,
\begin{gather}
\mathcal{P}_{\delta_1,\delta_2}^\mu=\delta_1 B_\alpha\, \delta_2\Big(\frac{\partial \mathcal{L}}{\partial(\partial_\mu B_\alpha)}\Big)+\delta_1 \rho\, \delta_2\Big(\frac{\partial \mathcal{L}}{\partial(\partial_\mu \rho)}\Big)+\delta_1 \phi\, \delta_2\Big(\frac{\partial \mathcal{L}}{\partial(\partial_\mu \phi)}\Big)-(1\leftrightarrow 2)\,,
\end{gather}
where $\delta_1$ and $\delta_2$ denote any two perturbations which solve the equations of motion \eqref{eq:eom}. Moreover, the asymptotic behaviour of the radial component gives,
\begin{align}\label{eq:scurrent_asymptotics}
P_{\delta_1,\delta_2}^r=&\frac{1}{r^3}\,\left(\delta_1\phi_{s}\,\delta_2\left(\sqrt{-\gamma}\,\langle\mathcal{O}_\phi\rangle\right)-\delta_2\varphi_{s}\,\delta_1\left(\sqrt{-\gamma}\,\langle\mathcal{O}_\phi\rangle\right)\right)\nn
&+\frac{1}{r^3}\,\left(\delta_1\rho_{s}\,\delta_2\left(\sqrt{-\gamma}\,\langle\mathcal{O}_\rho\rangle\right)-\delta_2\rho_{s}\,\delta_1\left(\sqrt{-\gamma}\,\langle\mathcal{O}_\rho\rangle\right)\right)\nn
&+\frac{1}{r^3}\,\left(\delta_1 m_a\,\delta_2\left(\sqrt{-\gamma}\,\langle J^a\rangle\right)-\delta_2 m_a\,\delta_1\left(\sqrt{-\gamma}\,\langle J^a\rangle\right)\right)+\cdots\,,
\end{align}
where we have used the expressions in equation \eqref{eq:vevs_radial_der} along with the fact that we work in the hydrodynamic limit. The latter allows us to drop the derivatives terms in the counterterms of equation \eqref{eq:bdy_action}.

The property which is crucial in our construction is the fact that for any two perturbations which solve the equations of motion \eqref{eq:eom}, the symplectic current density is divergence free,
\begin{equation}\label{sympl}
\partial_\mu\mathcal{ P}_{\delta_1,\delta_2}^\mu=0\,.
\end{equation}
We are going to construct two symplectic currents from our hydrodynamic and two static perturbations. The first one is $P_{\delta H,\delta_{\star}}$, which is made out of the static perturbation \eqref{btbp}. The second one is $P_{\delta H,\delta_{m_x}}$, made out of the static perturbation \eqref{ax}.

Given the fact that we can Fourier expand our modes, we find it convenient to do the same for the components of the symplectic current density according to,
\begin{align}
\mathcal{P}_{\delta_1,\delta_2}^\mu=e^{-i\omega (t+S(r))+i \ves q x}P_{\delta_1,\delta_2}^\mu\,.
\end{align}
The divergence free condition \eqref{sympl} gives,
\begin{gather}\label{symplv2}
-i\omega P^t_{\delta_1,\delta_2}+P^{r \prime}_{\delta_1,\delta_2}-i\omega S' P^r_{\delta_1,\delta_2}+i\ves q P^x_{\delta_1,\delta_2}=0\nn
\Rightarrow \left.P^r_{\delta_1,\delta_2}\right|_{r \to \infty}-\left.P^r_{\delta_1,\delta_2}\right|_{r \to 0}+\int^\infty_0 dr( -i\omega P^t_{\delta_1,\delta_2}-i\omega S^\prime P^r_{\delta_1,\delta_2}+i\ves q P^x_{\delta_1,\delta_2})=0\,.
\end{gather}
where in order to get the second line, we have integrated from the horizon to infinity.

Turning our attention to the specific examples for perturbations to be used in the symplectic current, we will first consider $\mathcal{P}_{\delta H,\delta_{\star}}$ which is made out of our hydrodynamic mode and the static perturbation in \eqref{backbp}. After performing an expansion of the Fourier components in $\ves$, we obtain,
\begin{align}
P^t_{\delta H,\delta_{\star}}=\mathcal{O}(\ves^2)\,,\qquad
P^x_{\delta H,\delta_{\star}}=-i\,q\,\delta a\, \delta \rho^2_{\star(0)}\,\ves+\mathcal{O}(\ves^2)\,,
\end{align}
and for the radial component we have,
\begin{gather}
P^r_{\delta H,\delta_{\star}}=-e^{2g}\Big(\delta a_{\star (2)}\delta_H (\tau W^{rt})_{[1]}-\delta \tilde{a}_{(2)}\delta_\star (\tau W^{rt})_{[1]}+\nn \frac{a U}{2}(\delta\rho_{\star(2)}\delta\rho'_{\star (0)}-\delta\rho'_{\star(2)}\delta\rho_{\star (0)})+ \frac{U}{2}(\delta\rho_{\star(0)}\delta\tilde{\rho}'_{ (2)}-\delta\rho'_{\star(0)}\delta\tilde{\rho}_{ (2)})+\nn U(\delta \phi_{\star(2)}\delta\tilde{\phi}'_{(2)}-\delta \phi'_{\star(2)}\delta\tilde{\phi}_{(2)})-i\omega_{[2]}S' U a \delta \rho_{\star(0)}^2\Big)\ves^2+\mathcal{O}(\ves^3)\,.
\end{gather}
Using the above relations for the symplectic current $\mathcal{P}_{\delta H,\delta_{\star}}$ in equation \eqref{symplv2} we obtain our first reduced equation,
\begin{gather}\label{dth0}
-\delta j_{t[1]}\delta\mu_{\star (2)}+i\omega_{[2]}\delta\varrho_{\star(2)}\delta\tilde{\theta}_v+(2\Delta_\phi-3)\delta\tilde{\phi}_{v(2)}\delta \phi_{s\star(2)}=\nn i\omega_{[2]}\delta \varrho_{h\star(2)}\delta\theta_0+i\omega_{[2]} \left(\delta\rho^{(0)}_{(0)\star}\right)^2 e^{2g^{(0)}}\,\delta a-q^2 \,\delta a \int^\infty_0 dr \delta \rho^2_{\star (0)}\,.
\end{gather}
As we will in the next section, the above relation will become the effective equation of motion for the amplitude of the order parameter.

We will now consider the symplectic current $\mathcal{P}_{\delta H,\delta_{a_x}}$ which is constructed from the hydrodynamic and the static perturbation of equation \eqref{btbp}. For the Fourier modes of the components along the field theory directions we have that,
\begin{align}
P_{\delta H,\delta_{a_x}}^t=\mathcal{O}(\ves^2)\,,\qquad  P_{\delta H,\delta_{a_x}}^x=\mathcal{O}(\ves^3)\,,
\end{align}
while for the radial component we have the non-trivial form,
\begin{gather}
P_{\delta H,\delta_{a_x}}^r=-U \tau_c \Big(\delta B^x_{x(0)}(-iq\delta \tilde{b}_{r(1)}+q\omega_{[2]}S'\delta\tilde{\theta}_v+\delta\tilde{b}'_{x(2)})-iq\delta\tilde{\theta}_v\delta B^{x \prime }_{x(2)}\Big)\ves^2+\mathcal{O}(\ves^3)\,.
\end{gather}
Substituting the asymptotic and near horizon expressions in equation \eqref{symplv2} at order $\ves^2$ we find that,
\begin{gather}\label{jx}
\delta j_{x(2)}=-i\chi_{jj}^{(2)}\,q\,\delta\tilde{\theta}_v+\tau^{(0)}_c q\omega_{[2]}\,(\delta \theta_0-\delta \tilde{\theta}_v)\,,
\end{gather}
which as we will later see, will yield a constitutive relation for the electric current along with the third line of equation \eqref{ids}.

\section{Effective Theory}\label{sec:effective}

In this section we will collect the results of section \ref{sec:deriv} to form our effective theory for a suitable set of hydrodynamic variables. We will do this in two different ways, using two different sets of variables. In the first approach, we will write a constitutive relation for the electric current in terms of the chemical potential and the amplitude and phase of the condensate. The effective theory will then be complete by fixing a Josephson relation and a first order time evolution equation for the amplitude of the order parameter. When combined with the Ward identity \eqref{eq:Ward}, we obtain a closed system of equations for the dynamics of the system. The second approach uses an effective energy potential which allows us to write the equations of motion for the charge density and the amplitude and phase of the condensate as a system of first order equations in time. This will allow us to compare with the Model F in the classification of Hohenberg and Halperin \cite{RevModPhys.49.435}.

\subsection{Hydro Description}

In order to clearly state our effective theory, we need to identify its dynamical variables. For this purpose, the most straightforward set of variables is the phase of the order parameter $\delta\tilde{\theta}_v$, the variable $\delta\theta_0$ and $\delta a$ which parametrises the amplitude of the order parameter according to,
\begin{align}
\langle\mathcal{O}_\psi\rangle=\Delta\langle\mathcal{O}_\psi\rangle_b\,\left(1+\ves^{-1}\,\delta a +i\,q_e\, \delta \theta_v\right)\,,
\end{align}
with $\Delta\langle\mathcal{O}_\psi\rangle=(2\,\Delta_\psi-3)\,\ves\,\delta\rho_{v\ast (0)}+\cdots$, the VEV of the complex scalar operator in the thermal state. For later convenience, we will define the new amplitude variable, $\delta\tilde{a}=\delta a/\ves$.

However, we still have the variation variable $\delta\mu_{\#}$ in our description which can be specified by the first line of \eqref{ids}. An alternative description, which seems more natural from the hydrodynamics point of view,  is to maintain the phase $\delta\tilde{\theta}_v$, the amplitude $\delta a$ and to trade $\delta\mu_{\#(2)}$ for the chemical potential variation defined by,
\begin{align}\label{eq:mu_def}
\delta\tilde{\mu}=\ves \left(\delta\mu_{\ast (2)}\,\delta a-\delta\mu_{\# (2)} \right)\,.
\end{align}
For any quantity $\mathcal{O}$ which is a function of the chemical potential $\mu$ and scalar deformation $\phi_s$, we can define the difference,
\begin{align}
\Delta\mathcal{O}=\mathcal{O}_\star(\mu_c+\delta\mu_\ast,\phi_s+\delta\phi_{s\ast})-\mathcal{O}_\#(\mu_c+\delta\mu_\ast,\phi_s+\delta\phi_{s\ast})\,.
\end{align}
This measures the difference of the value of $\mathcal{O}$ between the normal and the broken phase at fixed chemical potential and deformation parameter.

After these definitions, we can write the constitutive relations for the current,
\begin{align}\label{eq:constitutive_rels}
\delta J_t&=-\chi_{QQ}^\#\,\delta\tilde{\mu}-2\,\Delta\varrho\,\delta \tilde{a}\,,\nn
\delta J_i&=-\chi_{JJ}\,\partial_i\delta\theta_v-\sigma_d\,\partial_i\delta\tilde{\mu}\,,
\end{align}
which are nothing but equation \eqref{jx} and the third line of \eqref{ids} when combined with the first line of \eqref{ids}. In the above we have reinstated factors of $\ves$ and Fourier transformed back to a coordinate space description and we have also introduced the incoherent conductivity $\sigma_d=\tau^{(0)}$. Recasting equation \eqref{eqth} in terms of our new variables then provides a Josephson relation for the chemical potential variation $\delta\tilde{\mu}$,
\begin{align}
q_e^2 e^{2g^{(0)}}\left( \Delta \rho^{(0)}\right)^2\,\partial_t\delta\theta_v=\left(\chi_{QQ}^\#\,\partial_t -\sigma_d\,\partial_i\partial^i+q_e^2\,e^{2g^{(0)}}\left( \Delta \rho^{(0)}\right)^2\right)\,\delta\tilde{\mu}+2\,\Delta\varrho_h\,\partial_t\delta \tilde{a}\,.
\end{align}
By using the Ward identity \eqref{eq:Ward}, this equation can be written in the form of a Josephson relation for the local chemical potential,
\begin{align}\label{eq:josephson_dis}
\delta\tilde{\mu}=\partial_t\delta\theta_v-\frac{\chi_{JJ}}{q_e^2\,\varpi_1}\,\partial^2\delta\theta_v+\frac{\varpi_2}{q_e\,\varpi_1}\,\partial_t\delta \tilde{a}\,,
\end{align}
where we have defined the coefficients,
\begin{align}\label{eq:varpi_defs}
\varpi_1=\frac{s_c}{4\pi}\left( \Delta \rho^{(0)}\right)^2\,,\qquad \varpi_2=\frac{2}{q_e}\left(\Delta\varrho-\Delta\varrho_h \right)=\frac{2}{q_e}\,\left(\varrho_\ast-\varrho_{h\ast}\right)\,.
\end{align}
In the last equality we have used that in the normal phase the field theory charge density is equal to the horizon one and therefore $\varrho_{h\#}=\varrho_\#$.

From the above we see that in this notation our Josephson relation \eqref{eq:josephson_dis} contains dissipative effects. It is also evident that the amplitude degree of freedom has to enter both the constitutive relations \eqref{eq:constitutive_rels} as well as the Josephson relation \eqref{eq:josephson_dis}. The final equation we have left in order to have a complete description is \eqref{dth0} which fixed the dynamics of the amplitude. In our notation it reads,
\begin{align}\label{eq:amplitude_eom}
\varpi_1\,\partial_t\delta \tilde{a}-\left(8\,\Delta w_{FE}+\gamma\,\partial_i\partial^i \right)\delta \tilde{a}-q_e\,\varpi_2\,\partial_t\delta\theta_v-\left(2\,\Delta\varrho-q_e\,\varpi_2\right)\,\delta\tilde{\mu}=0\,,
\end{align}
with $\gamma$ as defined in \eqref{eq:gamma_def}. In order to obtain the equation above, we have used that the difference in the free energy density can be written as,
\begin{align}
\Delta w_{FE}=-\frac{1}{2}\left(\chi_{QQ}^\star-\chi_{QQ}^\#\right)\delta\mu_\ast^2-\frac{1}{2}\left(\nu_\phi^\star-\nu_\phi^\#\right)\delta\phi_{s\ast}^2-\left(\nu_\mu^\star-\nu_\mu^\#\right)\delta\phi_{s\ast}\,\delta\mu_{\ast}+\cdots\,,
\end{align}
at leading order in the variations $\delta\mu_\ast$ and $\delta\phi_{s\ast}$. This concludes the construction of our effective theory which is comprised of the equations \eqref{eq:josephson_dis}, \eqref{eq:amplitude_eom} and the Ward identity \eqref{eq:Ward} given the constitutive relations \eqref{eq:constitutive_rels}.

Finally, we would like to write the constitutive relation for the expectation value,
\begin{align}
\delta \langle\mathcal{O}_\phi\rangle=2\,\Delta\langle\mathcal{O}_\phi\rangle\,\delta\tilde{a}+\nu^\#_\mu\,\delta\tilde{\mu}\,,
\end{align}
 of the neutral scalar operator. The above relation follows from the last line of equation \eqref{ids} and the definitions \eqref{eq:suscepts_norm} and \eqref{eq:mu_def}.

\subsection{Matching with model F}
In this subsection we will compare the effective theory we finalised in the previous subsection to the Model F of Hohenberg and Halperin \cite{RevModPhys.49.435}. In order to do this, we will need to rewrite our theory in terms of the amplitude $\delta a$, the angle $\delta\theta_v$ and the charge density $\delta\tilde{\varrho}=\delta J^t$ fluctuations. To do this we can simply invert the constitutive relation for the time component of the electric current in equation \eqref{eq:constitutive_rels}. After solving for the time derivatives of the fields in our description by using equations \eqref{eq:josephson_dis}, \eqref{eq:amplitude_eom} and \eqref{eq:Ward} we have the system of first order equations in time,
\begin{align}\label{eq:modelf_eoms}
\partial_t \delta \tilde{\varrho}&=\frac{\sigma_d}{\chi^\#_{QQ}}\, \partial_i\partial^i(\delta\tilde{\varrho}-2\, \Delta\varrho\,\delta \tilde{a})+\chi_{JJ}\, \partial_i\partial^i \delta\theta_v\,,\nn
\partial_t \delta \tilde{a}&= \lambda_1 \left(\gamma\, \partial_i\partial^i \delta \tilde{a}+8\,\left.\Delta E\right|_{\varrho,\phi_s}\delta \tilde{a}+\frac{2\,\Delta \varrho}{\chi^\#_{QQ}}\,\delta \tilde{\varrho}\right)+\frac{\chi_{JJ}}{q_e}\, \lambda_2\, \partial_i\partial^i \delta\theta_v\,,\nn
\partial_t \delta\theta_v&=\frac{\chi_{JJ}}{q_e^2} \lambda_1 \,\partial_i\partial^i\delta\theta_v+\frac{1}{\chi^\#_{QQ}}\left(\delta \tilde{\varrho}-2\,\Delta\varrho\,\delta \tilde{a}\right)\nn
&\qquad- \frac{\lambda_2}{q_e}\left( \gamma\,\partial_i\partial^i \delta \tilde{a}+8\,\left.\Delta E\right|_{\varrho,\phi_s}\delta \tilde{a}+\frac{2\,\Delta \varrho}{\chi^\#_{QQ}}\delta\tilde{\varrho}\right) \,.
\end{align}
In order to simplify the notation we have introduced the quantity,
\begin{align}\label{eq:energy_def}
\left.\Delta E\right|_{\varrho,\phi_s}=\left.\Delta w_{FE}\right|_{\mu,\phi_s}-\frac{1}{2\chi_{QQ}^\#}\,\left.\left(\Delta\varrho\right)^2\right|_{\mu,\phi_s}\,,
\end{align}
which is the energy density difference of the broken and the normal phase at fixed charge density and scalar deformation. The above relation is easy to show by e.g. using the results of Appendix B in \cite{Donos:2022xfd}. Moreover, in equations \eqref{eq:modelf_eoms}, we have defined the two important quantities,
\begin{align}\label{eq:lambda_def}
\lambda_i=\frac{\varpi_i}{\varpi_1^2+\varpi_2^2}\,,.
\end{align}

We now need to match the above system of equations to the equations that one would obtain from the phenomenologically motivated equations of the Model F of \cite{RevModPhys.49.435}. In order to do that, we first need to consider the Ginzburg-Landau free energy functional close to the critical point,
\begin{align}\label{eq:GL_free_energy}
F[\psi,m]=\int d^2 x \left( \frac{w_0}{2} | \nabla \psi |^2+\frac{\tilde{r_0}}{2}| \psi |^2+\tilde{u_0} | \psi |^4+\frac{1}{2\,C_0} m^2+\gamma_0\, m \,| \psi|^2 \right)\,,
\end{align}
where in this notation $m$ is the conserved charge density. Given the above energy functional, the dissipative equations of motion for the $U(1)$ order parameter $\psi$ and the current continuity equation respectively are,
\begin{align}\label{HHeqs}
\partial_t \psi &=-2\Gamma_0 \frac{\delta F}{\delta \psi^\star}-ig_0 \psi\, \frac{\delta F}{\delta m}\,,\nn
\partial_t m &=\lambda_0^m\, \nabla^2 \frac{\partial F}{\partial m}+2\,g_0\, \mathrm{Im}(\psi^\star \frac{\delta F}{\delta \psi^\star})\,.
\end{align}
After decomposing the order parameter as $\psi=\Delta\langle\mathcal{O}_\psi\rangle\,\left(1+\delta \tilde{a}-i\,g_0\,\delta\theta_v\right)$ and the charge density as $m=m_0+\delta\tilde{\varrho}$, we can match the resulting equations of motion provided that,
\begin{align}\label{eq:match}
g_0&=q_e\,,\qquad C_0=\chi^\#_{qq}\,,\qquad \gamma_0=-\frac{\Delta\varrho}{\chi^\#_{QQ}\,\left(\Delta\langle\mathcal{O}_\psi\rangle\right)^2}\,,\qquad w_0=\frac{\chi_{JJ}}{q_e^2\, \left(\Delta\langle\mathcal{O}_\psi\rangle\right)^2}\,,\nn
m_0&=\Delta\varrho\,,\qquad \tilde{u}_0=-\frac{1}{\left(\Delta\langle\mathcal{O}_\psi\rangle\right)^4}\,\left.\Delta E\right|_{\varrho,\phi_s}\,,\qquad \tilde{r}_0=4\,\frac{\Delta w_{FE}}{\left(\Delta\langle\mathcal{O}_\psi\rangle\right)^2}\,,\nn
\Gamma_0&=\left(\Delta\langle\mathcal{O}_\psi\rangle\right)^2\,\left(\lambda_1+i\,\lambda_2 \right)\,,\qquad \lambda_0^m=\sigma_d=\tau^{(0)}\,.
\end{align}
The important lesson that we extract from having explicit expressions for these constants from holography is that apart from $\tilde{r}_0$ and $m_0$, the rest remain finite as we approach the critical point. Notably, the dissipative kinetic coefficient $\Gamma_0$ remains complex as $\ves\to 0$. These observations will play an important role in the next subsection where we discuss the hydrodynamic modes as we approach the critical point from both the normal and the broken phases. 

It is important to note that the coefficients \eqref{eq:match} are ultimately fixed by information which is held fixed either due to conservation laws, like the charge density $\varrho$, or because it is part of the sources in the problem, like the deformation parameter $\phi_{s}$.  We have written quantities, like the susceptibilities $\chi_{QQ}^{\#}$, $\chi_{QQ}^\star$ and the free energy $w_{FE}$ which are more natural for the grand canonical ensemble. However, these are to be evaluated on states of chemical potential which is specified by the fixed charge density $\varrho$.

\section{Hydrodynamic Modes}\label{sec:modes}

In this section we will consider the quasinormal modes of the system which is captured by the effective theory that we constructed in section \ref{sec:effective}. More specifically, we will be interested in three different regimes of the phase space while holding fixed the wavevector of fluctuations $k_i$. The first one will be the normal phase as we approach the critical point. 

The two subsequent subsections are devoted to two distinct limits of the general, finite density case, for small and for large values of the gap as compared to the modulus of the wavevector. As we will see, at finite density the limits of approaching the transition from the normal and the broken phase give a discontinuity in the dispersion relations. In other words, at a fixed wavevector the dispersion relations of the our quasinormal modes are discontinuous across the phase transition. The final regime we will examine is the broken phase at zero chemical potential. In this case, we will have a great simplification of the dispersion relations and we will be able to follow the modes in the complex plane analytically. Moreover, this section will answer some of the questions raised in \cite{Donos:2021pkk}.

\subsection{Normal Phase}
In this subsection we will consider the hydrodynamic modes of our system as we approach the critical point from the normal phase of the system. In general, at exactly the critical point the only gapless modes of our system are fluctuations of the charge density and the critical modes of the complex scalar that become gapless. Even though we are at finite charge density, the fact that we are in the probe limit suggests that the charge density fluctuations will decouple from the pressure and the momentum of system leading to a purely diffusive mode.

More specifically, in the normal phase the constitutive relations for the fluctuations of the electric current are,
\begin{align}
\delta J^t=\chi_{QQ}^\#\,\delta\mu\,,\qquad \delta J^i=-\sigma_d\,\partial^i\delta\mu\,.
\end{align}
From a holographic point of view, the incoherent conductivity is given by $\sigma_d=\tau^{(0)}$ in the normal phase. Imposing the Ward identity \eqref{eq:Ward} yields a mode with dispersion relation,
\begin{align}\label{eq:el_dif}
\omega=-i\,\frac{\sigma_d}{\chi_{QQ}^\#}\,k^2=-i\,\frac{\tau^{(0)}}{\chi_{QQ}^\#}\,k^2\,.
\end{align}

In order to understand the mode relevant to the charged scalar, we will employ once again the symplectic current of our theory. Suppose that $\delta\rho_{(0)}$ is the static mode at the critical point. In order to construct the finite wavevector one we perturbatively expand it in the wavevector according to,\
\begin{align}
\delta\rho_H&=e^{-i\omega (t+S(r))+i \ves q x}\,\left(\delta\rho_{(0)}+\ves\,\delta\rho_{(1)}+\ves^2\,\delta\rho_{(2)}+\cdots \right)\,,\nn
\omega&=\ves\,\omega_{[1]}+\ves^2\,\omega_{[2]}+\cdots\,.
\end{align}
By considering the symplectic current that we can form from the static mode $\delta\rho_{(0)}$ and the above hydrodynamic expansion, it is easy to show that the dispersion relation of the critical mode is diffusive with,
\begin{align}
\omega=-i\,\frac{4\pi}{s_c}\frac{1}{\left(\delta\rho_{(0)}^{(0)}\right)^2}\,\int_0^\infty dr\,\left(\delta\rho_{(0)}\right)^2\,k^2\,.
\end{align}
Even though our computation was purely in the normal phase, it will be useful to write this as a limit coming from the broken phase after observing that,
\begin{align}
\frac{1}{\left(\delta\rho_{(0)}^{(0)}\right)^2}\,\int_0^\infty dr\,\left(\delta\rho_{(0)}\right)^2=\lim_{\ves\to 0}\frac{\gamma}{\left(\rho^{(0)}\right)^2}\,,
\end{align}
yielding,
\begin{align}\label{eq:op_dif}
\omega=-i \frac{4\pi}{s_c}\,k^2\,\lim_{\ves\to 0}\frac{\gamma}{\left(\rho^{(0)}\right)^2}=-i \frac{4\pi}{q_e^2\,s_c}\,k^2\,\lim_{\ves\to 0}\frac{\chi_{JJ}}{\left(\rho^{(0)}\right)^2}\,,
\end{align}
where we have used the relation \eqref{eq:gamma_chi} which holds close to the phase transition.

In fact, we don't only have one such diffusive modes but two. This comes from the fact that above the critical temperature we can decompose the complex scalar into a real and a complex part. Each one of those satisfies exactly the same equation of motion at a perturbative level. The above analysis shows that right above the critical temperature we have three diffusive modes.

\subsection{Large Gap Limit}\label{subsec:Large_Gap}

In this subsection we will consider the limit in which the module of the wavevector is much smaller than the expected gap of the amplitude mode. Since we are working in a probe limit,  the expectation is that we will recover the regular hydrodynamics of the supercurrent. This should happen after integrating out the amplitude mode which will acquire a large gap. Before doing that we will examine what happens with the modes of our effective theory in the large gap limit.

In order to do this we perform a Fourier decomposition of our modes and recast the linearised equations of motion \eqref{HHeqs} in matrix form,
\begin{align}\label{eq:disp_rel}
\mathbb{M}\left(\omega,k_i\right)\,
\begin{pmatrix}
\delta a_0\\ \delta \theta_{v0}\\ \delta\tilde{\varrho}_0
\end{pmatrix}
=0\,.
\end{align}
In order for the perturbation to admit non-trivial solutions, the matrix $\mathbb{M}$ must be non-invertible and it should therefore have zero determinant. This condition becomes a third order algebraic equation for $\omega$, that fixes the dispersion relations $\omega(k_i)$ of our quasinormal modes. 

The general solution of this system is quite complicated but we will consider the two limits of small and large wavevectors in this and the next subsections. For small wavevectors, we perform the expansion,
\begin{align}
k_i&=\lambda\,q_i\,,\nn
\omega&=\omega_{[0]}+\lambda\,\omega_{[1]}+\lambda^2\,\omega_{[2]}+\cdots\,,
\end{align}
and solve the equation order by order in $\lambda$. As one would expect, the three modes we find consist of two sound and one gapped diffusive mode. The first few terms in a wavevector modulus expansion are,
\begin{align}\label{eq:sound_mode}
\omega_H&=i\,\mathrm{Re}\Gamma_0\,\frac{8\,\left.\Delta E\right|_{\varrho,\phi_s}}{\left(\Delta\langle\mathcal{O}_\psi\rangle\right)^2}\nn
&\qquad -i\left(w_0\,\mathrm{Re}\Gamma_0-\frac{\left(\Delta\varrho\right)^2\,\lambda_0^m}{2\left(\chi_{QQ}^\#\right)^2\,\left.\Delta E\right|_{\varrho,\phi_s}}-\frac{w_0}{\mathrm{Re}\Gamma_0}\left(\mathrm{Im}\Gamma_0+\frac{q_e\,\Delta\varrho \left(\Delta\langle\mathcal{O}_\psi\rangle\right)^2}{4\,\chi_{QQ}^\#\left.\Delta E\right|_{\varrho,\phi_s}} \right)^2 \right)k^2\,,\nn
\omega_{\pm}&=\pm \sqrt{\frac{\chi_{JJ}}{\chi_{QQ}^\star}\,k^2}\nn
&\qquad -\frac{i}{2}\left(w_0\,\mathrm{Re}\Gamma_0+ \frac{\lambda_0^m}{\chi_{QQ}^\star}+\frac{w_0}{\mathrm{Re}\Gamma_0}\left(\mathrm{Im}\Gamma_0+\frac{q_e\,\Delta\varrho \left(\Delta\langle\mathcal{O}_\psi\rangle\right)^2}{4\,\chi_{QQ}^\#\left.\Delta E\right|_{\varrho,\phi_s}} \right)^2 \right)k^2\,.
\end{align}
In order to obtain the above result, we have used the non-obvious relation,
\begin{align}\label{eq:susc_rel}
\left.\Delta E\right|_{\varrho,\phi_s}=\frac{\chi^\star_{QQ}}{\chi^\#_{QQ}}\,\Delta w_{FE}\,,
\end{align}
which we prove in Appendix \ref{app:suscept}.

It is reassuring to note that the gap of the first dispersion relation, corresponding to the  Higgs/Amplitude mode, agrees with the expression of \cite{Donos:2022xfd} even though we are only in the probe limit. Moreover, the diffusion constant of the same mode is not positive definite and it remains finite as we approach the critical point. Finally, as we will see in subsection \ref{sec:zero_mu}. in more detail, its limiting value agrees with the diffusion constant of \eqref{eq:op_dif} only in the case of zero charge density.

It is nice to see that the modes behave as we would expect them to in our limit. However, it is important to understand how to reduce the theory of section \ref{sec:effective} to regular hydrodynamics and reproduce the two sound modes that we derived from the full theory. An appropriate limit to take is,
\begin{align}
\partial_t\to\lambda\,\partial_t\,,\quad \partial_i\to\lambda\,\partial_i\,,\quad \delta a\to \lambda\,\delta a\,,\quad \delta\tilde{\mu}\to\lambda\,\delta\tilde{\mu}\,,
\end{align}
with $\lambda$ a small dimensionless parameter. By doing this we can find the local expression for the amplitude,
\begin{align}
\delta \tilde{a}=-\frac{1}{8\,\Delta w_{FE}}\left( 2\,\Delta\varrho\,\delta\tilde{\mu}+\frac{\chi_{JJ}}{q_e^2\,\lambda_1}\left(q_e \lambda_2+\frac{q_e^2\,\Delta\varrho}{4\,\Delta w_{FE}\,\chi_{QQ}^\star}\right)\,\partial^2\theta_v\right)\,,
\end{align}
where we have included corrections up to order $\mathcal{O}(\lambda^2)$. Moreover, the Josephson relation for the redefined chemical potential $\delta\hat{\mu}$ becomes,
\begin{align}\label{eq:josephson_regular}
\delta\hat{\mu}&=\partial_t\delta\theta_v-\chi_{JJ}\,\zeta_3\,\partial^2\delta\theta_v\nn
&=\partial_t\delta\theta_v-w_0\,\left(\mathrm{Re}\Gamma_0+\frac{1}{\mathrm{Re}\Gamma_0} \left( \mathrm{Im}\Gamma_0+\frac{q_e\,\Delta\varrho\,\left(\Delta\langle\mathcal{O}_\psi\rangle\right)^2}{4\,\Delta w_{FE}\,\chi_{QQ}^\star}\right)^2\right)\,\partial^2\delta\theta_v\,,
\end{align}
where we have also used the definition of the third bulk viscosity $\zeta_3$ for a superfluid. This is true since the new chemical potential $\delta\hat{\mu}$ is chosen so that we are in the transverse frame with,
\begin{align}
\delta J_t&=-\chi_{QQ}^\star\,\delta\hat{\mu}\,,\nn
\delta J_i&=-\chi_{JJ}\,\partial_i\delta\theta_v-\lambda_0^m\,\partial_i\,\delta\hat{\mu}\,.
\end{align}
Given the above, the resulting dispersion relation,
\begin{align}
\omega_{\pm}=\pm\sqrt{\frac{\chi_{JJ}}{\chi_{QQ}^\ast}\,k^2}-\frac{i}{2\,\chi_{QQ}^\star}\left( \lambda_0^m+\chi_{QQ}^\star\,\chi_{JJ}\,\zeta_3\right)\,,
\end{align}
matches equation \eqref{eq:sound_mode}. Finally, it is worth noting that the expression for the bulk viscosity $\zeta_3$ in \eqref{eq:josephson_regular} agrees with the one we recently reported in \cite{Donos:2022www} when we take its limit near criticality.

\subsection{Small Gap Limit}\label{subsec:Small_Gap}
In this subsection we would like to understand the behaviour of our modes in the limit where we keep the wavevectors fixed but we approach the critical point from the side of the broken phase. In this limit, the gap becomes small or equivalently, the wavevector is large. The equations of our effective theory become,
\begin{align}
\partial_t\delta\tilde{\varrho}&=\frac{\lambda_0^m}{\chi_{QQ}^\#}\,\partial_i\partial^i\delta\tilde{\varrho}\,,\nn
\partial_t\delta \tilde{a}&=w_0\,\left(\mathrm{Re}\Gamma_0\,\partial_i\partial^i\delta \tilde{a}+q_e\,\mathrm{Im}\Gamma_0\,\partial_i\partial^i\,\delta\theta_v\right)+\frac{2\,\mathrm{Re}\Gamma_0\,\Delta\varrho}{\chi_{QQ}^\# \left(\Delta\langle\mathcal{O}_\psi\rangle\right)^2}\,\delta\tilde{\varrho}\,,\nn
\partial_t\delta\theta_v&=w_0\,\left(\mathrm{Re}\Gamma_0\,\partial_i\partial^i\delta\theta_v-\frac{1}{q_e} \mathrm{Im}\Gamma_0\,\partial_i\partial^i\delta \tilde{a}\right)+\frac{1}{\chi_{QQ}^\#}\left(1-\frac{2\,\mathrm{Im}\Gamma_0\,\Delta\varrho}{q_e \left(\Delta\langle\mathcal{O}_\psi\rangle\right)^2} \right)\,\delta\tilde{\varrho}\,.
\end{align}

The Jacobi form of the system immediately gives that there is a charge diffusion modes with dispersion relation,
\begin{align}\label{eq:higgs_largek}
\omega=-i \frac{\lambda_0^m}{\chi_{QQ}^\#}\,k^2\,,
\end{align}
which matches exactly the charge diffusion mode of equation \eqref{eq:el_dif} of the normal phase given the matching of the parameters of equation \eqref{eq:match}. It is worth noting that apart from the charge density, the phase and the amplitude will also be involved in this mode in the broken phase. This is true even in the limit close to the critical point since the constants $\Gamma_0$ and $w_0$ remain finite.

The second mode involves only the order parameter as in order to satisfy the first equation of the system we have to necessarily set the charge density fluctuation equal to zero. A quick computation shows that the order parameter fluctuations yields two modes obeying the dispersion relations,
\begin{align}\label{eq:gold_largek}
\omega_r=-i\,w_0\,\Gamma_0\,k^2\,,\qquad \omega_r=-i\,w_0\,\bar{\Gamma}_0\,k^2\,,
\end{align}
which remain finite close to the critical point. It is worth comparing with the diffusive modes \eqref{eq:op_dif} of the order parameter that we find in the normal phase. Using the matching of equation \eqref{eq:match}, and the definitions \eqref{eq:lambda_def} we see that this pair of modes matches the modes of oscillation of the order parameter in the normal phase \eqref{eq:op_dif}, only at zero chemical potential. At finite density the dispersion relations are discontinuous since $\Gamma_0$ remains complex and finite in the near critical limit.

\subsection{Zero chemical potential}\label{sec:zero_mu}

It is easy to see that at zero chemical potential,  the matching conditions \eqref{eq:match} suggest that,
\begin{align}
\mathrm{Im}\Gamma_0=\gamma_0=\rho_c=0\,.
\end{align}
In this case, the equation of motion for the amplitude of the order parameter is not sourced by fluctuations of the phase and the charge density leading to the pseudo-gapped mode with dispersion relation,
\begin{align}\label{eq:Higgs_neutral}
\omega_H=i\,\Gamma_0\,\left(\frac{8\,\left.\Delta E\right|_{\varrho,\phi_s}}{\left(\Delta\langle\mathcal{O}_\psi\rangle\right)^2}-w_0\,k^2\right)\,.
\end{align}
Note that this dispersion relation agrees exactly with the dispersion \eqref{eq:op_dif} of the complex scalar coming from the normal phase in the limit of zero gap. This is reassuring but we are still missing one more mode with the same diffusion constant. As we will see, this will come from the sector of charge density and phase.

In order to find the quasinormal modes of this sector, it is illustrating to write down their equations of motion,
\begin{align}
\partial_t \delta \tilde{\varrho}&=\frac{\lambda_0^m}{\chi_{QQ}^\#}\,\partial_i\partial^i\delta\tilde{\varrho}+\chi_{JJ}\, \partial_i\partial^i \delta\theta_v\,,\nn
\partial_t \delta\theta_v&=w_0\,\Gamma_0 \,\partial_i\partial^i\delta\theta_v +\frac{1}{\chi^\#_{QQ}}\delta \tilde{\varrho}\,,
\end{align}
and note once again that the amplitude decouples entirely. The corresponding dispersion relations are,
\begin{align}\label{eq:Gold_neutral}
\omega_{\pm}=\pm \frac{\sqrt{4\,k^2\,\chi_{JJ}\,\chi_{QQ}^\#-k^4\,\left(\lambda_0^m-w_0\,\Gamma_0\, \chi_{QQ}^\#\right)^2}}{2\chi_{QQ}^\#}-i\frac{\lambda_0^m+w_0\,\Gamma_0\,\chi_{QQ}^\#}{2\,\chi_{QQ}^\#}\,k^2\,.
\end{align}
The limit we want to examine is the small $k$ limit in which we find the approximate dispersion relation,
\begin{align}
\omega_{\pm}\approx \pm \sqrt{\frac{\chi_{JJ}}{\chi_{QQ}^\#}k^2}-i\frac{\lambda_0^m+w_0\,\Gamma_0\,\chi_{QQ}^\#}{2\chi_{QQ}^\#}\,k^2\,,
\end{align}
which agrees with the dispersion relations of \cite{Donos:2021pkk} when taking the nearly critical limit. These are therefore the standard sounds modes of neutral superfluids in the broken phase.

The second point we would like to make comes from thinking of this dispersion relation as the position of poles of Green's functions in the complex frequency plane. Using our analytic formula \eqref{eq:Gold_neutral} we see that the two sound-like poles collide on the imaginary axis when,
\begin{align}\label{eq:k_crit}
k_c^2=\frac{4\,\chi_{JJ}\,\chi_{QQ}^\#}{\left(\lambda_0^m-w_0\mathrm{Re}\Gamma_0 \chi_{QQ}^\#\right)^2}\,.
\end{align}
We also see that for wavevectors with modulus squared larger than $k_c^2$, the two modes remain purely imaginary.

For very large value of $k^2$ as compared to $k_c$, we have the two approximate dispersion relations with leading behaviour,
\begin{align}
\omega_\pm=-\frac{i}{2\chi_{QQ}^\#}\left(\lambda_0^m+w_0\,\mathrm{Re}\Gamma_0\,\chi_{QQ}^\#\pm \left|\lambda_0^m-w_0\,\mathrm{Re}\Gamma_0\,\chi_{QQ}^\# \right|\right)\,k^2\,.
\end{align}
The above shows that the collision of the two sound modes produce the, so far missing, second diffusive mode that matches \eqref{eq:op_dif} as well as the charge diffusion mode \eqref{eq:el_dif}.

\section{Numerical checks}\label{sec:numerical_checks}

In this section we check numerically the dispersion relations for the Higgs and Goldstone modes close to the critical point for various values of the wavevector. First, we give an overview of our method and then we focus on two simple cases: (i) Charged superfluids at finite density and zero scalar deformation parameter $\phi_s$ and (ii) Neutral superfluids with a non-trivial scalar deformation parameter $\phi_s$.

\subsection{Overview of the method}
In this subsection we give a few technical details on the double sided shooting method we have used. First, we describe the technique for the background solutions and then we move on to discuss the quasi-normal modes we wish to construct.

\subsubsection{Background solution}

We are going to work in the probe limit, assuming that the matter fields don't backreact onto the metric, which will be of the general form \eqref{eq:background_metric} with asymptotics  described by the expressions  \eqref{eq:metric_uv},\eqref{eq:metric_ir}. As in the analytic calculation, the conformal boundary is located at $r \to \infty $ and the black hole horizon at $r=0$.

The matter action is taken to be \eqref{eq:bulk_actionv1} with:
\begin{align}
V&=\frac{1}{2} m_{\psi}^2\, |\psi|^2+\frac{1}{2} m_\phi^2\, \phi^2+\lambda_\psi \, |\psi|^4+\lambda_\phi\, \phi^4+\lambda_{\psi \phi}\,|\psi|^2 \phi^2\,,\nn
\tau&=1+\zeta_\psi\, |\psi|^2+\zeta_\phi\, \phi^2\,.
\end{align}

A background matter solution describing both the broken and normal phase of the system involves an ansatz of the form,
\begin{align}
B=a(r) dt\,,\qquad \rho=\rho(r)\,,\qquad \phi=\phi(r)\,.
\end{align}
Plugging this ansatz in the equations of motion we find 3 second order ODEs, which implies that we need to fix six integration constants. The behaviour of the fields near the conformal boundary is,
\begin{align}
a&=\mu-\frac{\varrho}{(r+R)}+\cdots\,,\nn
\rho&=\frac{\rho_v}{(r+R)^{\Delta_\psi}}+\cdots\,,\nn
\phi&=\frac{\phi_s}{(r+R)^{3-\Delta_\phi}}+\cdots+\frac{\phi_v}{(r+R)^{\Delta_\phi}}+\cdots\,.
\end{align}
Close to the horizon, the analytic expansion yields,
\begin{align}
a=a^{(0)} r+\cdots\,,\quad \rho=\rho^{(0)}+\cdots\,,\quad \phi=\phi^{(0)}+\cdots\,.
\end{align}

Fixing the values of the chemical potential $\mu$ and the neutral scalar's source $\phi_s$ leaves six free integration constants  $ \varrho$, $\rho_v$, $\phi_v,a^{(0)}$,  $\rho^{(0)},\phi^{(0)} $ that will be fixed via double-sided shooting. For an appropriate choice of the parameters $\lambda_\psi,\lambda_\phi,\lambda_{\psi \phi},\zeta_\psi,\zeta_\phi$ we can find background solutions describing a second order phase transition. As expected, the field $\rho$ is going to be trivial in the normal phase of the system and non-trivial in the broken phase.

With the background solutions at hand, we can  construct  the functions $\varrho(\mu,\phi_s)$ and $\phi_v(\mu,\phi_s)$ numerically. By taking the appropriate partial derivatives of these functions we can calculate the static susceptibilities $\chi_{QQ},\nu_\phi,\nu_\mu$ that appear in our analytic results. To calculate the current susceptibility $\chi_{JJ}$ we need to construct a static perturbation for the one-form field of the form
\begin{align}
\delta B =\delta b_x (r)\, dx\,,
\end{align}
yielding a second order ODE for the function $\delta b_x(r)$ which requires fixing of two constants of integration in order to find a unique solution. This is the bulk dual of a field theory perturbation involving the supercurrent.The expansion near the conformal boundary at $r\to\infty$ is,
\begin{equation}
\delta b_x =\delta b_x^s+\frac{\delta b_x^v}{r+R}+\cdots\,.
\end{equation}
Close to the black hole horizon we have the analytic expansion,
\begin{equation}
\delta b_x =\delta b_x^{(0)} +\cdots\,.
\end{equation}

In total we have three free integration constants ($\delta b_x^s,\delta b_x^v,\delta b_x^{(0)}$) and because the equation for  $\delta b_x$ is linear and homogeneous we can set one of them to unity. This argument shows that there are indeed two free constants of integration and we can find a unique solution. Then we are able to find $\chi_{JJ}$ using the definition: 
\begin{equation}
\chi_{JJ}=-\frac{\delta b_x^v}{\delta b_x^s}\,.
\end{equation}
At this point, we remind the reader that this static perturbation is part of our black hole thermodynamics. However, we only consider perturbations of it as we wish to study the hydrodynamics of superfluid thermal states with zero supercurrent.

\subsubsection{Quasinormal modes}

Our ultimate goal is to calculate the dispersion relations $\omega (k)$ of the hydrodynamic modes and for this reason  we need to study black hole perturbations which are source free from the boundary point of view. To achieve this, we consider perturbations of the form,
\begin{equation}\label{func dep}
 \delta\mathcal{F}(r,t,x)=e^{-i\omega( t+S(r))+i k x}\delta f(r)\,,
\end{equation}
with the choice,
\begin{equation}
S(r)=\int^r_\infty \frac{dy}{U(y)}\,.
\end{equation}

The longitudinal sector that we are interested in involves the fields $\delta b_t,\delta b_x, \delta b_r, \delta \rho,\delta \phi$.  The radial component equation for the gauge field allows us to eliminate $\delta b_r$ in terms of $\delta b_t$ and $\delta b_x$. This leaves us with four second order ODEs which require the fixing of eight constants of integration. Once again, we will achieve this by implementing a double-sided shooting method.

In order to find identify the constants of integration, we consider the asymptotic behaviour of our functions close to the boundaries of our computational domain. In the IR we impose in-falling boundary conditions and solving the equations of motion we find the expansions,
\begin{align}
\delta b_t=c_t+\cdots\,,\quad
\delta b_x=c_x+\cdots\,,\quad
\delta \rho=c_\rho+\cdots\,,\quad
\delta \phi=c_\phi +\cdots\,,
\end{align}
where the constants $c_t,c_x,c_\rho,c_\phi$ are unfixed at this stage. On the other side of our domain, in the UV, we have the expansions,
\begin{align}
\delta b_t&=\delta b_t^s-i \omega \delta c+\frac{\delta b_t^v}{r+R}+\cdots\,,\nn
\delta b_x&=\delta b_x^s+i k \delta c+\frac{\delta b_x^v}{r+R}+\cdots\,,\nn
\delta \rho&=\frac{\delta \rho_s}{(r+R)^{3-\Delta_\psi}}+\cdots+\frac{\delta \rho_v}{(r+R)^{\Delta_\psi}}+\cdots\,,\nn
\delta \phi&=\frac{\delta \phi_s}{(r+R)^{3-\Delta_\phi}}+\cdots+\frac{\delta \phi_v}{(r+R)^{\Delta_\phi}}+\cdots\,,
\end{align}
where the constant $\delta b_x^v$ is fixed in terms of the others, due to the current conservation \eqref{eq:Ward}.

In order to compute the quasinormal modes we have to set the sources to zero,
\begin{align}
\delta \rho_s=\delta \phi_s=\delta b_t^s=\delta b_x^s=0\,.
\end{align}
In total, for fixed a fixed value of $k$, we have 4 independent constants from the IR ($c_t,c_x,c_\rho,c_\phi$) and 4 independent constants from the UV ($\delta b_t^v,\delta \rho_v,\delta \phi_v, \delta c$). Because the equations of the perturbations are linear and homogeneous we can set one of those constants to unity, so we are left with 7 free constants plus the frequency $\omega$, which match exactly the 8 constants of integration that we need.

\subsection{Results for charged superfluids}

In this model we take the metric to be AdS-Schwarzschild with unit radius. In this case, the background geometry \eqref{eq:background_metric} is specified by the functions,
\begin{align}\label{bgr}
U(r)=(r+R)^2-\frac{R^3}{r+R}\,,\qquad
g(r)=\log(r+R)\,.
\end{align}
which give Hawking temperature is given by $T=\frac{3R}{4 \pi}$ and entropy density $s=4\pi R^2$. For our charged superfluids, we have chosen to set the neutral scalar's background source $\phi_s$ to zero which allows us to consistently set the bulk scalar $\phi$ equal to zero in our equations of motion. Moreover, we will choose $m_\psi^2=-2 , \zeta_\psi=0, \lambda_\psi=0$ and also set $T=\frac{3}{4\pi}$, without loss of generality. For these parameters the system exhibits a second order phase transition with critical chemical potential $\mu_c \approx 4.06371366  $.

\begin{figure}[h!]
\centering
\includegraphics{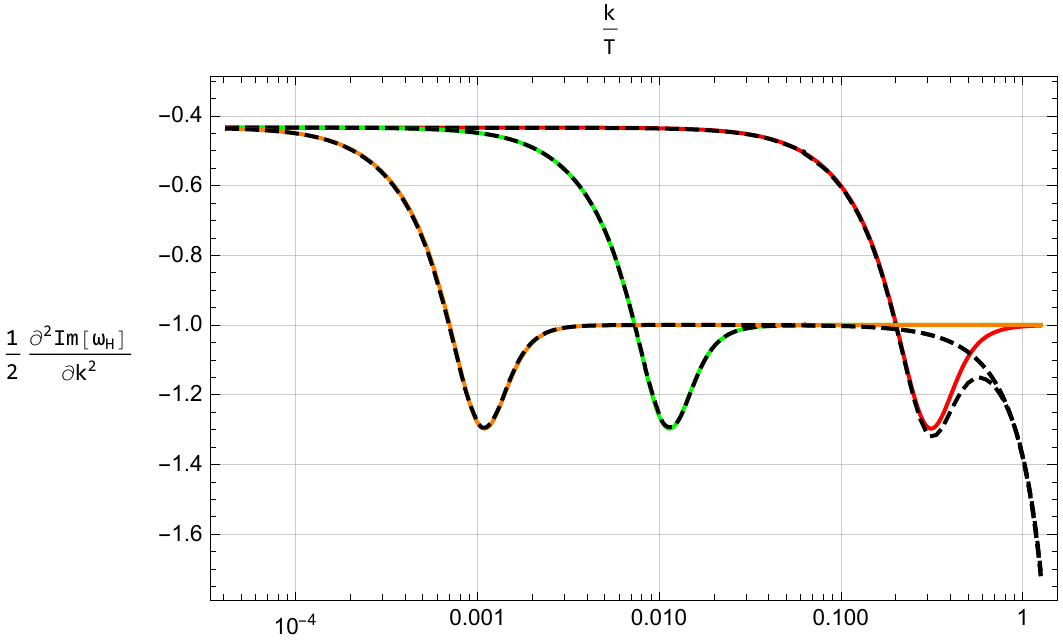}
\caption{Plot of $\frac{1}{2}\frac{\partial^2 \mathrm{Im}\,\omega_H}{\partial k^2}$ for the Higgs mode as a function of k for three values of the chemical potential $\frac{\mu_c}{\mu_{orange}}=0.99999999,\frac{\mu_c}{\mu_{green}}=0.999998932,\frac{\mu_c}{\mu_{red}}=0.9991919498$. The dashed lines correspond to the numerical results and the solid lines to the analytic predictions.}
\label{fig:higgs_charged}
\end{figure}

In Figure  \ref{fig:higgs_charged} we plot the quantity $\frac{1}{2}\frac{\partial^2 \mathrm{Im}\,\omega_H}{\partial k^2}$ for the Higgs mode as a function of $k$, for three different values of the chemical potential above $\mu_c$. We plot the results coming from the numerical calculation  (dashed lines) together with the analytic results (solid lines) that we can find for the frequency as a function of the wavevector. The latter is one of the roots of the cubic polynomial resulting from demanding that \eqref{eq:disp_rel} has non-trivial solutions.

As we can observe, sufficiently close to the critical point  the Higgs mode interpolates between two regions where it is diffusive, in accordance with the discussion in sections \ref{subsec:Large_Gap} and \ref{subsec:Small_Gap}. In particular, for $k$ much smaller than the  gap, the first equation of \eqref{eq:sound_mode}  yields for our model: $\omega_H\approx \omega_{gap}-0.43374\,i\, k^2$ and for $k$ much greater than the gap equation \eqref{eq:higgs_largek} gives $\omega_H\approx-i \,k^2$.

The authors of \cite{Amado:2009ts}, working with exactly the same setup, argued that this  mode is diffusive with $\omega=\omega_{gap}-i\, k^2$. As we showed here, this is indeed the correct behaviour but only for values of momentum much larger than the gap. The reason they didn't find the interpolation  is that their numerical calculation was done  only for k larger than the gap, as we can see from Figure 9 of their paper.

\begin{figure}[h!]
\centering
\includegraphics[width=0.48\linewidth]{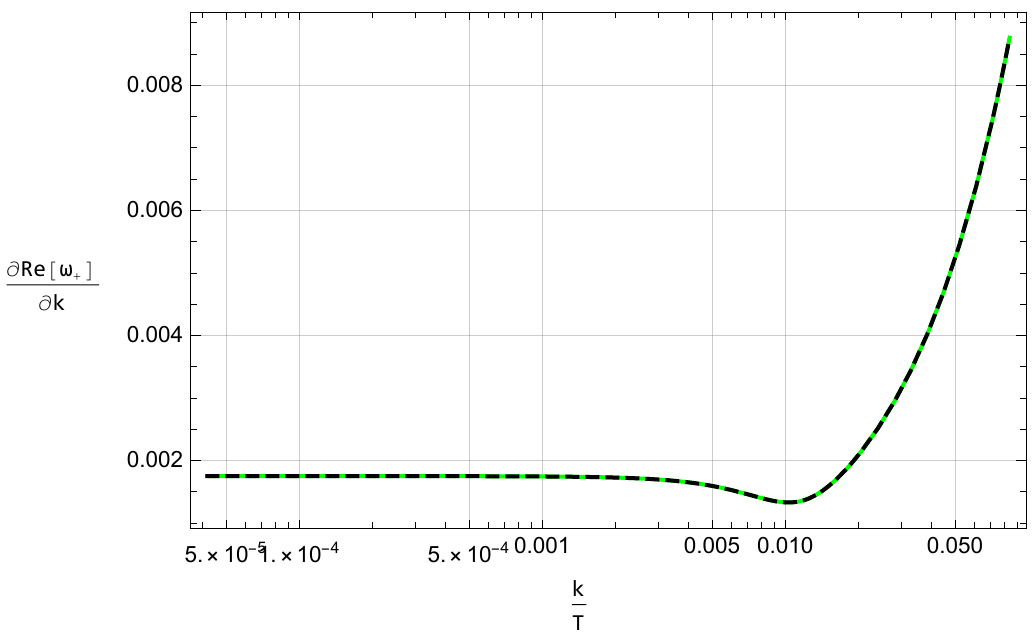}\quad\includegraphics[width=0.48\linewidth]{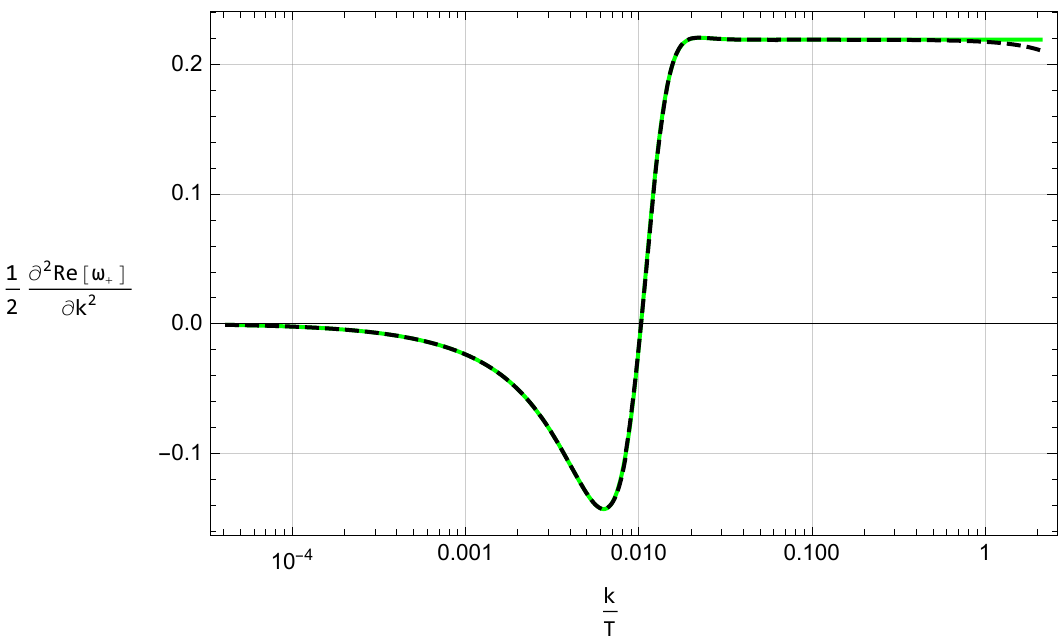}\\
\includegraphics[width=0.48\linewidth]{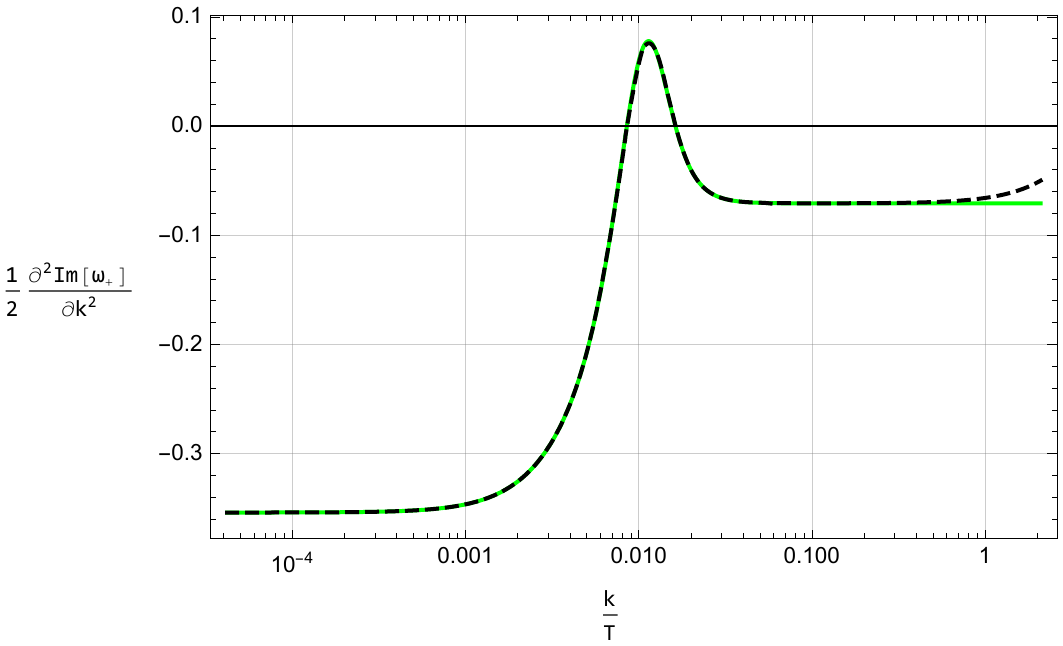}
\caption{Plots of $\frac{\partial \mathrm{Re}\,\omega_{+}}{\partial k}$,  $\frac{1}{2}\frac{\partial^2\mathrm{Re}\, \omega_{+}}{\partial k^2}$ and $\frac{1}{2}\frac{\partial^2 \mathrm{Im}\,\omega_{+}}{\partial k^2}$ for the Goldstone mode as a function of k for  $\frac{\mu_c}{\mu}=0.999998932$. The dashed lines correspond to the numerical results and the solid lines to the analytic predictions.}
\label{fig:Gold_charged}
\end{figure}

In Figure \ref{fig:Gold_charged}  we plot $\frac{\partial \mathrm{Re}\omega_{+}}{\partial k}$,  $\frac{1}{2}\frac{\partial^2\mathrm{Re} \omega_{+}}{\partial k^2}$ and $\frac{1}{2}\frac{\partial^2 \mathrm{Im}\omega_{+}}{\partial k^2}$ for the Goldstone mode as a function of k for  $\frac{\mu_c}{\mu}=0.999998932$.   We present  the results coming from the numerical calculation  (dashed lines) together with the analytic results (solid lines). The latter is the second root of the cubic that also fixed the dispersion relation of the Higgs mode. Apart from showing the agreement with the full solution, these plots also confirm the asymptotic behaviour given by equations \eqref{eq:sound_mode} and  \eqref{eq:gold_largek}. In particular these tell us that for k much smaller than the gap the dispersion relation approaches the behaviour $\omega_{+}\approx  0.001748 k -0.3541\,i\, k^2$ while for $k$ much larger than the gap: $\omega_{+}\approx \left(0.218982 -0.07098\,i\right) k^2 $.  Overall, for both modes there is a very good quantitative agreement with the analytic predictions, for all values of $\frac{k}{T} \ll 1$.

\subsection{Results for neutral superfluids}

Here we will study the quasi-normal modes of neutral superfluids which undergo a phase transition driven by the scalar deformation parameter $\phi_{(s)}$. We will one again choose the background metric \eqref{eq:background_metric} to be AdS-Schwarzschild with the functions given by \eqref{bgr} and Hawking temperature  $T=\frac{3}{4\pi}$. For the backgrounds we will set $\mu=0$ and we will also choose $m_\psi^2=-2$, $m_\phi^2=-2$, $\zeta_\psi=1$, $\zeta_\phi=1$, $\lambda_\psi=\frac{1}{2}$, $\lambda_\phi=\frac{1}{2}$ and $\lambda_{\psi \phi}=-\frac{3}{2}$. For this choice of parameters we find a second order phase transition with the critical value of the neutral scalar source being $\phi_{(s) c} \approx 2.5646887676$. An important observation is that for neutral superfluids we have two decoupled sectors, namely $(\delta b_t,\delta b_x)$ and $ (\delta \rho, \delta \phi)$. As one might expect, the first sector will capture the two quasinormal modes with relevant to the Goldstone mode while the second sector will capture the gapped Higgs mode. This is in accordance with the discussion in section \ref{sec:zero_mu}

\begin{figure}[h!]
\centering
\includegraphics{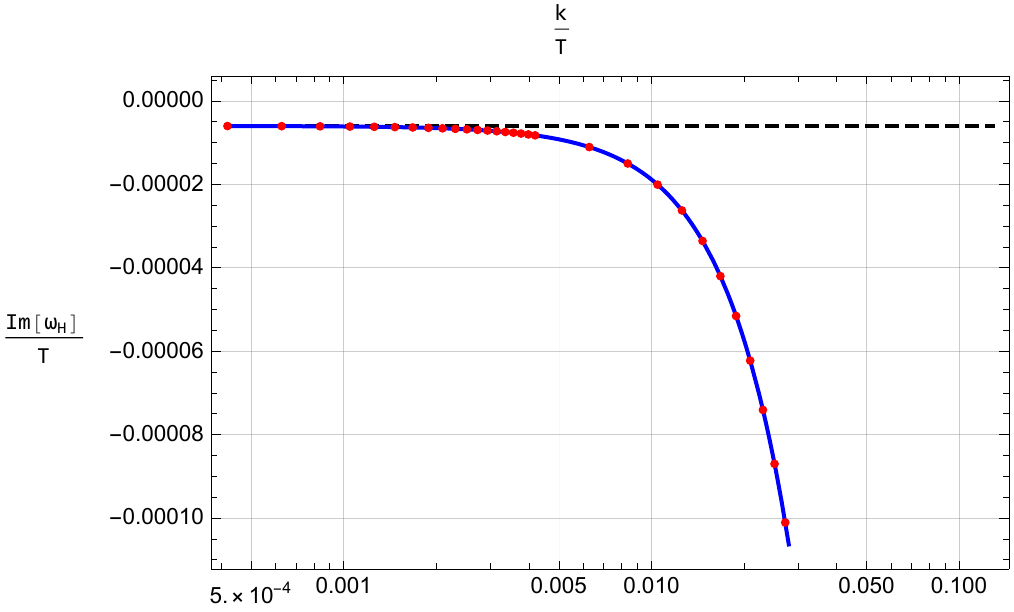}
\caption{Plot of $\frac{\omega_{H}}{T}$ as a function of k for $\frac{\phi_{(s) c}}{\phi_{(s)}}=0.999999519$. The dots are the numerical results and the solid line is the analytic result. The dashed horizontal line marks the analytic prediction for the gap of the Higgs mode.}
\label{fig:higgs_neutral}
\end{figure}

In Figure  \ref{fig:higgs_neutral} we plot our numerical results for the Higgs mode $\omega_H$ as a function of k for $\frac{\phi_{(s) c}}{\phi_{(s)}}=0.999999519$. In addition to that, we plot the corresponding analytic result that we find using equation \eqref{eq:Higgs_neutral}. The dashed horizontal line   shows the gap of the Higgs mode, which is $\frac{\omega_{gap}}{T}\approx-6.01\cdot 10^{-6} \,i$.

\begin{figure}[h!]
\centering
\includegraphics[width=0.48\linewidth]{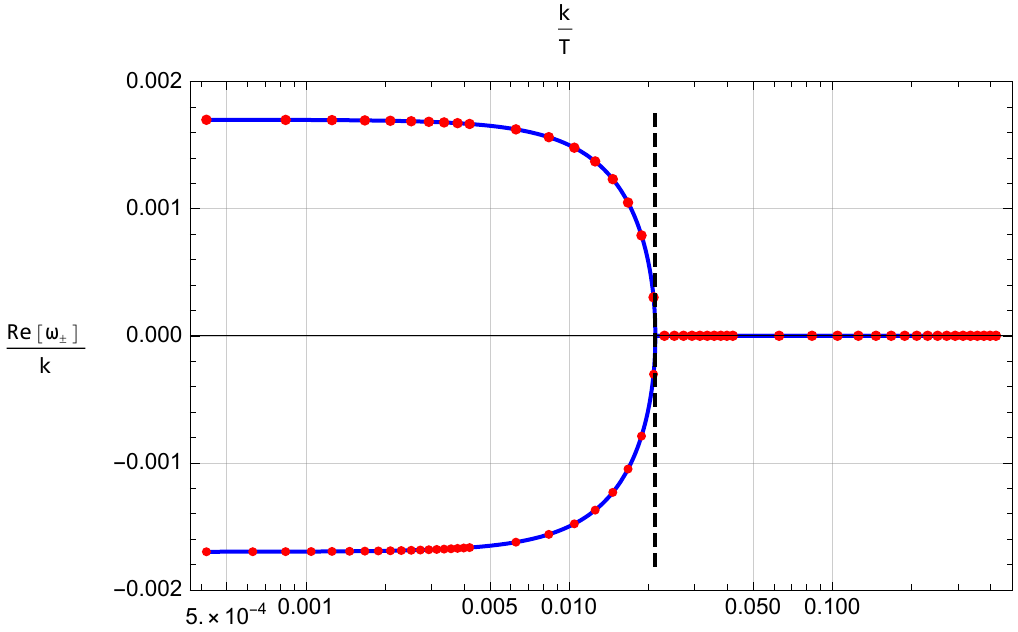}\quad\includegraphics[width=0.48\linewidth]{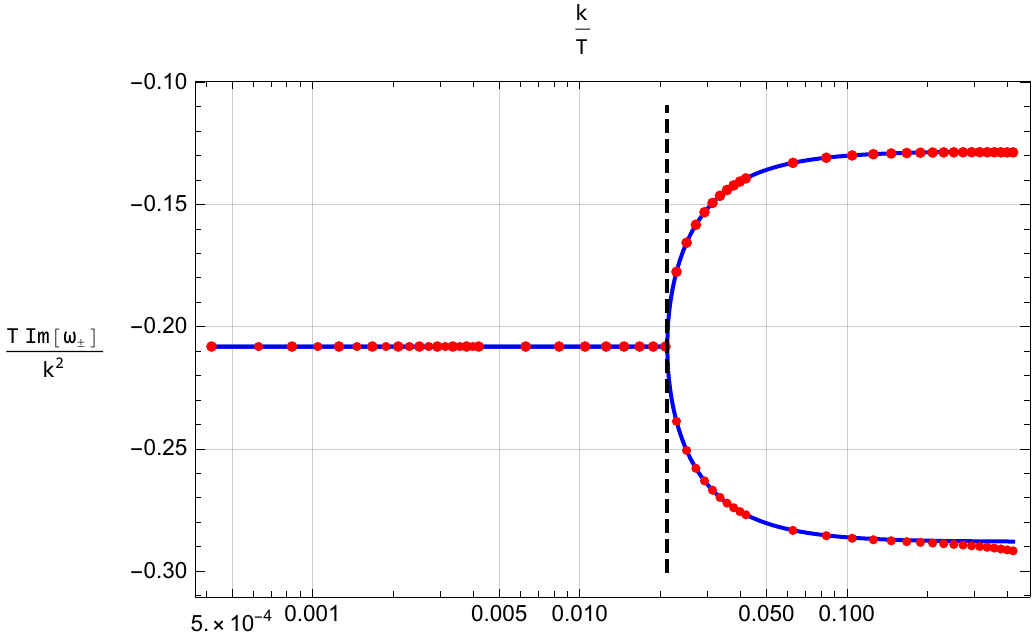}
\caption{Plots of $\frac{Re\omega_{\pm}}{k}$ and $\frac{Im\omega_{\pm}}{k^2}$ as a function of k for $\frac{\phi_{(s) c}}{\phi_{(s)}}=0.9999996754$. The dots are the numerical  results and the solid lines are the analytic predictions. The dashed lines are at  $\frac{k}{T}=0.0212842$.}
\label{fig:Gold_neutral}
\end{figure}

For the Goldstone mode,  we plot $\frac{\mathrm{Re}\,\omega_{\pm}}{k}$ and $\frac{T\, \mathrm{Im}\,\omega_{\pm}}{k^2}$,  as a function of $k$ for $\frac{\phi_{(s) c}}{\phi_{(s)}}=0.9999996754$ in Figure \ref{fig:Gold_neutral}. We also include plots of the analytic predictions (solid lines) of equation \eqref{eq:Gold_neutral}. The dashed lines on these plots  mark the point of collision for the two Goldstone modes at momentum $\frac{k_c}{T}=0.0212842$, as computed from equation \eqref{eq:k_crit}. These plots clearly show that for $\frac{k}{k_c}\ll 1$  we  can observe the usual second sound modes of superfluids, whereas for $\frac{k}{k_c} \gg 1$ these modes become purely diffusive.  As we explained in section \ref{sec:zero_mu} one of these two diffusive modes pairs nicely with the Higgs mode of Figure \ref{fig:higgs_neutral} agreeing with the doublet or modes of the charged scalar in the normal phase. Once again we find very good quantitative agreement with our analytic predictions.

\section{Discussion and outlook}\label{sec:discussion}

In this paper we have analysed the low energy dynamics of holographic superfluids close to their critical point. As part of our analysis, we have constructed an effective theory for the collective degrees of freedom in involved in the problem. For the standard description of superfluids away from the critical point, the hydrodynamic degrees of freedom are captured by the phase of the order parameter and the local chemical potential \cite{PhysRev.60.356,PhysRev.72.838}. Close to the critical point, the amplitude mode driving the transition becomes gapless and has to be included in the description.

By using analytic techniques developed in e.g. \cite{Donos:2022www,Donos:2022uea} we have constructed an effective consisting of the conservation law \eqref{eq:Ward}, the Josephson relation \eqref{eq:josephson_dis} and the equation governing the amplitude dynamics \eqref{eq:amplitude_eom}. By performing a change of variables,  we managed to show that our system is equivalent to Model F in the classification of Hohenberg and Halperin \cite{RevModPhys.49.435}. This allowed us to give explicit expressions for the various constants that appear in that description including the dissipative kinetic coefficient. With this information in hand, we carried out a somewhat detailed analysis of our modes in different limits for the gap of the Higgs mode. Moreover, given the analytic understanding of the modes in the normal phase,  we have revealed interesting discontinuities in their dispersion relations.

Our work can be extended in several different directions. An obvious direction would be to enlarge our minimal description to accommodate for the temperature and fluid velocity of the normal component of our system. In order to do this, we would need to move away from the probe limit and include the backreaction of the background metric along with the coupling of the complex scalar one-form field fluctuations with the metric. In such a more complicated scenario our techniques would still produce usable results. In a sense that would our recent analysis \cite{Donos:2022www} would need to be extended with the inclusion of the Higgs mode that we discussed in the present simplifying case.

\section*{Acknowledgements}

We would like to thank M. Baggioli and V. Ziogas for discussions on the topic. We would also like to thank the organisers of the workshop "Recent Developments in Strongly-Correlated Quantum Matter" for providing a stimulating environment where an important part of this work was carried out. AD is supported by STFC grant ST/T000708/1.

\appendix

\section{Susceptibility relations}\label{app:suscept}

The aim of this appendix is to prove equation \eqref{eq:susc_rel}. In order to do this, we remind the reader a couple of facts about the free energy difference between the normal and broken phases $\Delta w_{FE}(\mu,\phi_s)$.  Since we don't vary temperature in our probe model, we have suppressed the dependence on it. The basic property that this function satisfies is that it vanishes on the critical hypersurface $\left(\mu_c(\phi_s),\phi_s \right)$,
\begin{align}
\Delta w_{FE} \left(\mu_c(\phi_s),\phi_s \right)=0\,.
\end{align}
Moreover, with the transition being second order, the normal derivative with respect to the hypersurface also vanishes. These two statement imply that we indeed have,
\begin{align}
\nabla\Delta w_{FE}\left(\mu_c(\phi_s),\phi_s \right)=0\,,
\end{align}
and that the points $\left(\mu_c(\phi_s),\phi_s \right)$ are not extrema.

This shows that the Hessian matrix of $\Delta w_{FE}(\mu,\phi_s)$ evaluated on the critical surface should only have one non-zero eigenvalue which should also be positive. As a consequence, the determinant of the Hessian should vanish showing the relation,
\begin{align}
\left( \nu_\mu^\star-\nu_\mu^\#\right)^2=\left( \nu_\phi^\star-\nu_\phi^\#\right)\,\left(\chi^\star_{QQ}-\chi^\#_{QQ}\right)\,.
\end{align}
Using this relation, it is then easy to show that,
\begin{align}
\Delta w_{FE}=-\frac{1}{\chi^\star_{QQ}-\chi^\#_{QQ}}\frac{\left(\Delta\varrho\right)^2}{2}\,,
\end{align}
which is equivalent to \eqref{eq:susc_rel} given equation the relation \eqref{eq:energy_def}.
\newpage
\bibliographystyle{utphys}
\bibliography{refs}{}
\end{document}